# "Non-Classical Crystal Growth Recipe" using nanocrystalline ceria: a detailed review.

S.K. Padhi [a,†], M. Ghanashyam Krishna [a,b]

*a. School of Physics and Advanced Centre of Research in High Energy Materials, University of Hyderabad, Hyderabad 500046, India. †Email:* ***spadhee1@gmail.com***

*b. Centre for Advanced Studies in Electronics Science and Technology, School of Physics, University of Hyderabad, Prof C R Rao Road, Hyderabad 500046, Telangana, India.*

# "Non-Classical Crystal Growth Recipe" using nanocrystalline ceria: a detailed review.

## Graphical Abstract

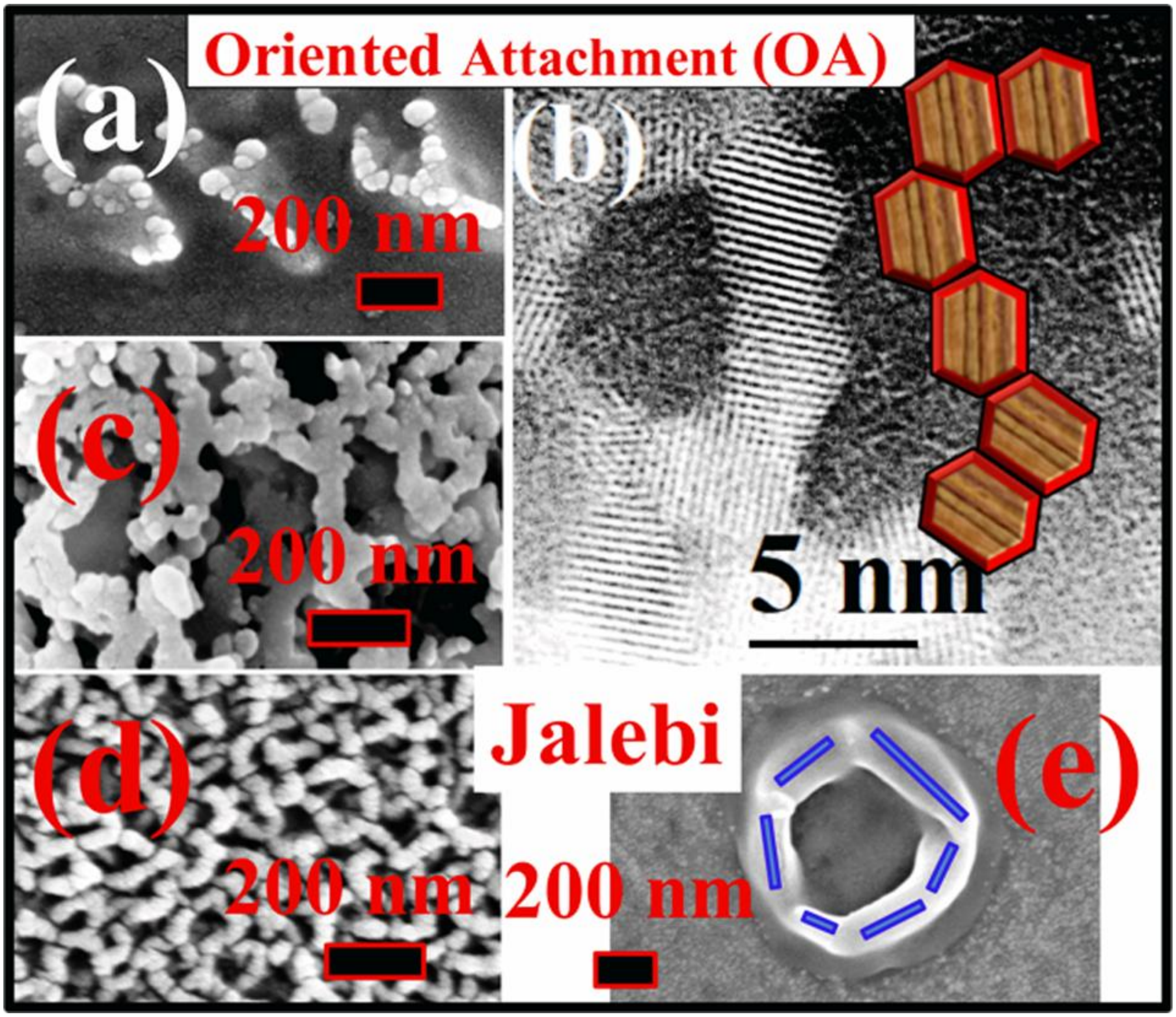

# Abstract

In this review, room temperature (RT) precipitation of the nanocrystalline-ceria (nc-ceria) re-dispersed and subsequently size-reduced by 20 kHz probe sonication in 25 % ethylene glycol/ 75 % DI-water mixed media is investigated. The sonication result in three nanostructured products: (1) water-soluble supernatant nc-ceria (Ce_Sl@RT), (2) settled gelatinous nc-ceria mass (Ce_SS@RT), and (3) ambient dried nc-ceria solid powder (Ce_SP@RT) product along with the parent RT nc-ceria (Ce@RT) precipitates. Surface/interface attributes are investigated systematically with the help of suitable spectroscopic probes. By following this synthesis protocol, the nc-ceria is made to cohabit with a variety (water, ethylene glycol, air) of neighbors that lead to the distinct surface and interface termination. The physical and chemical aspects of these varieties of the specialized surface terminated nc-ceria are explored coherently with respect to the Ce@RT precipitate. The second aspect of this review is devoted to the biomineralization for which the sonication derived Ce_Sl@RT is the candidate of choice. Aging of Ce_Sl@RT is physically tracked to mimic the natural aquatic medium crystal growth by the biomineralization process. In-situ TEM is extensively used to demonstrate the non-classical crystal growth mechanism physically. Uniquely TEM electron beam (e-beam) is exploited to aid both in the material manipulation and probing.

## Introduction:

Non-classical crystal growth (NCG) is a nature formulated enriched protocol recently explored to grow advanced hierarchical materials of complex morphologies and composition [1–10]. NCG briefly is the "***crystallization by building-units attachment***." The building units' attachment is brought about by either material-specific or neighboring environment directly contributing to the interaction and is the recognized sole growth deciding factor. Thus, in this inorganic crystal growth pathway the rule is to feasibly control interaction of growth directing building units' in order to realize tunable and novel technologically relevant morphologies.

There are numerous reports on (A) ***building units***, (B) ***growth driving force***, and (C) ***the final evolved hierarchical morphologies*** following the NCG protocol. Examples of ***building units shape*** are (1) PbSe nanocrystals [8], (2) ZnO nanocrystals [11], (3) α-$MnO_2$ nanowires [12], (4) 10-20 nm diameter $PbTiO_3$ fibers [13], (5) Au spherical seeds [14], (6) 1D-$TiO_2$ fibers [15], (7) δ-$MnO_2$ nanosheets [16], (8) 200-250 nm $SnO_2$ dodecahedrons [1]. The examples for ***corresponding growth directing forces*** are (1) dipolar interaction [8], (2) direction specific ZnO-ZnO interfacial interaction [11], (3) $K^+$-stabilized ionic interaction [12], (4) surface electrostatic force [13], (5) intense laser pulses [14], (6) electric field [15], (7) hydrogen bonding [16], (8) pressure induced [1]. While for the ***subsequent developed hierarchical morphologies*** are (1) PbSe nanowires and nanorings [8], (2) 1D-ZnO nanorods [11], (3) tunneled α-$MnO_2$ nanowires [12], (4) large-size 3D-$PbTiO_3$ hollowed fiber [13], (5) triangular Au nanoplates [14], (6) multilevel twinned tree-like branched $TiO_2$ [15], (7) nanoflowers [16], (8) μ-sized 3D-ordered $SnO_2$ superstructures [1], respectively. In addition, in our research lab hydrothermal 1D-ZnO growth observations in the context of NCG as a model system has been explored. The following sequence is (A) ***building units shape***: shown as top-view having (1) nanoparticles, (2) needles, (3) 3D-hexagonal discs, and (4) 2D-hexagonal plates; whereas the (C) ***corresponding evolved morphologies***: as side-view (1)-(4) of 1D- grown ZnO are illustrated in figs.1.0 (a)-(d) respectively.

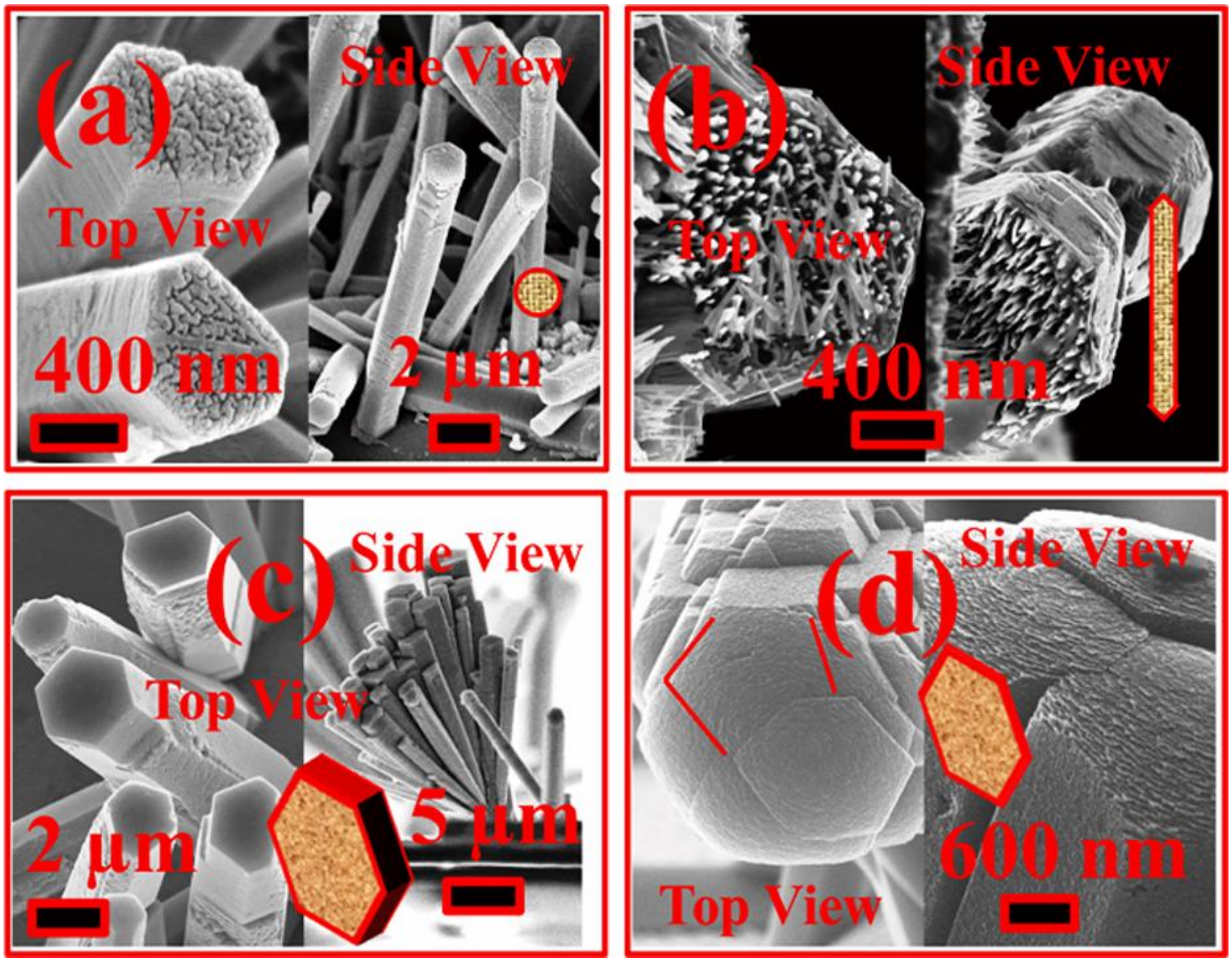

**Fig.1.0** [***Hydrothermally grown 1D-ZnO validating NCG scheme***]: Building units in (a)-(d) as side view respectively (courtesy Y. Rajesh et al. unpublished data).

Importantly the unique morphology of aquatic biominerals unique morphology and hierarchical motifs is the motivation and basis of the current NCG investigation [17]. In fresh and salt water μ-sized silica (spongy skeleton in soft bodies) and calcium carbonate (appropriate mechanical strength at the sea bed living) are two ubiquitous biominerals macrostructures and microstructures as shown in figs.1.1 (a)-(f) respectively [18]. The recent activities involving studies on biominerals formation and its extension to mimic a variety of other inorganic materials for hierarchical morphology development is ongoing [1,8,11–16,19–23].

Significantly, even the role of water as an active participant is investigated in bringing localized order for inducing attachment [11,24–28].

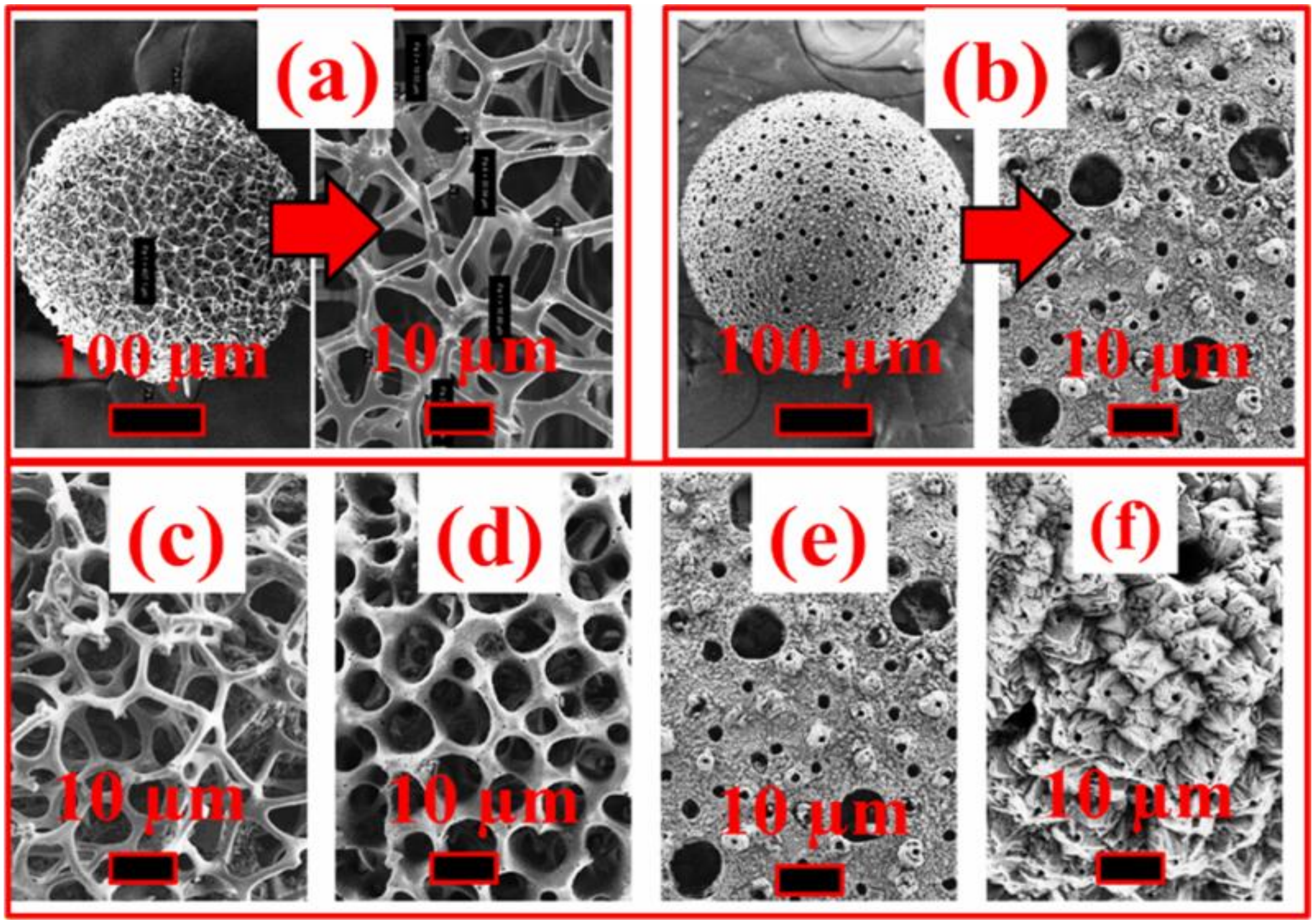

**Fig.1.1** [***Aquatic biominerals***]: (a) radiolarian siliceous, (b) foraminifera calcareous skeleton, and (c)-(f) are microstructural evolution adopted to sustain neighbouring aquatic environment respectively (courtesy Prof. A. C Narayana unpublished data).

The building unit's attachment in systematic fashion by oriented attachment (OA), delivers an intermediate structural feature identified as mesocrystal [29–32]. The R Penn et al. TEM illustration [33] and continual exploration of OA during the ongoing attachment process is constantly under investigation to develop new avenues. Besides ongoing application of the OA in providing novel materials [34–38], most recent publication such as "OA Revisited: Does a Chemical Reaction Occur ?" add valuable insights to the existing concepts [39]. Although mesocrystalline ceria had plenty OA demonstrations [40,41], this is not true in case of nanocrystalline ceria (nc-ceria). An investigation relating to the use of $Ce^{3+}$ surface charge steric difference with that of the grain interior $Ce^{4+}$ in aqueous medium by hydrothermal processing is of significance to present study [42]. Illustration of the aqueous hydrothermal processing developed mesocrystalline 1D-ceria feature from the nc-ceria particles OA process is shown in figs.1.2 (a) - (d). In it individual nc-ceria retains its microstructural distinctness but as a whole behaves as a single crystalline.

In realization of NCG for demonstration of the ongoing nanoscale physical ordering an essential part is played by the Transmission Electron Microscopy (TEM). The importance of such can be easily adjudged from the TEM micrographs presented in figs.1.2 (a) - (d), in which a reaction quenched but not completely crystallized 1D-ceria intermediate reaction product localized structure is shown. These depict OA of nc-ceria particles. It is a usual ex-situ TEM observation employed for NCG validation. However, frequent technique is used to dynamically track the NCG process is in-situ liquid cell TEM/STEM [43–48]. In general, TEM electron beam (TEM e-beam) is employed to probe. In contrast, its utilization in material modification in a controlled way is more limited [49–51]. Most recently Asghar et al. investigated nc-ceria justifying TEM as a standalone technique to probe and modify [50]. In this study TEM e-beam activates radiolysis of water and water acts as a direct participant in achieving NCG growth of 1D-ceria.

Based on the literature reviewed on NCG, this review aims to formulate and demonstrate in entirety, the NCG process as a recipe. The nc-ceria is chosen as the prototypical material for evaluation because of its unique attributes, such as (1) proven versatility for biological applications, (2) cyclic oxidation-reduction, and (3) radiation damage injection and self-healing by bringing a change to its immediate/near neighbour environment. These surface attributes are re-evaluated in the present context specific to the role of aqueous oxidized and reduced neighbour. The process of biomineralization is mimicked by designing nc-ceria water soluble colloidal dispersion. Lastly, it is demonstrated that the TEM e-beam can be explicitly employed as an in-situ probe as well as enabler of NCG. This particular case study of technologically important ceria can be extended to similar other radiation sensitive materials having also the radiation healing capacity for localized in-situ nanomaterials processing.

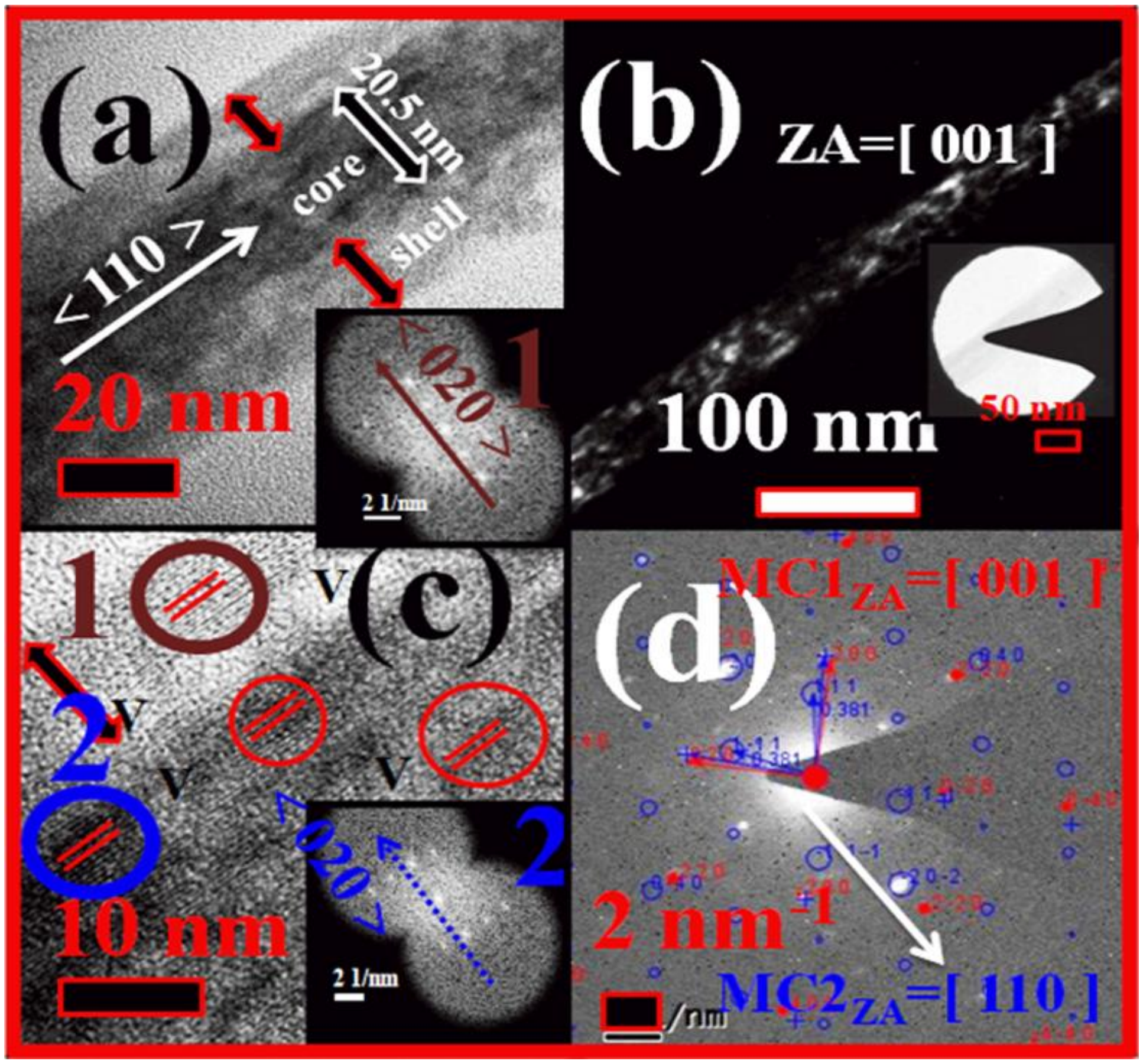

**Fig.1.2** [***1D-ceria Mesocrystal***]: TEM (a) BF distinct core-shell region, (b) DF individual crystals as brighter dots, (c) FFT of nc-ceria crystal one from the core and other from shell region identical spot pattern with no rotation (shell crystal-1 inset in fig.1.2 (a), core crystal-2 as inset in fig.1.2 (c)), and (d) TEM-SAED spot pattern from the entire region showing lattice mismatch of core and shell an justification to OA character.

## Synthesis and Charge Transfer (CT) transition:

## Synthesis Protocol:

A RT precipitation-redispersion strategy is employed to deliver ultrafine and DI-water dispersible nc-ceria. The formulated protocol based on literature review is presented below [52–56]. 100 mM ammonium cerium (IV) nitrate (Sigma-Aldrich, 99.99 % trace metal basis) precursor dispersed in 25 % ethylene glycol (Finar reagent, 99 % assay)/ 75 % di-water mixed media as the solvent is prepared in a round bottom flask. To these a 10 mL of 25 % ammonia solution (Finar, 0.91 dense) maintained at 400 rpm stirring label is introduced rapidly. A pH~10 is achieved, resulting in the precipitation. The stirring is continued for 2 hrs and the precipitate is allowed to settle overnight. The solid product is separated by centrifugation and is allowed to dry at ambient laboratory conditions over a month. This product, in the rest of the review, will be referred to as Ce@RT. Re-dispersion in DI-water is carried out with the use of the 13 mm solid-ultrasonic horn (Sonics VCX 750W, 20 kHz) operated at its 50 % amplitude label. In the second case, the centrifuged 5 g precipitate is placed in the bottom of a 250 mL standard sonochemical reaction vessel having 150 mL DI-water as a solvent for the re-dispersion. The sonication time is fixed for 30 minutes. The sonication prepared water-soluble supernatant liquid, the sol settled at the vessel bottom obtained after overnight storage, and also the ambient dried solid recovered out of the settled sol are taken for detailed spectroscopic investigations. During the sonication, ice-water bath is maintained around the sonochemical reaction vessel to assimilate the sonication generated thermal effect. The sonication derived supernatant liquid (Sl), semi-solid sol (SS), and the semi-solid sol converted into solid power (SP) product will be subsequently coded as Ce_Sl@RT, Ce_SS@RT, and Ce_SP@RT respectively in the rest of the review. The experimental elucidation of these three distinct (Ce_Sl@RT, Ce_SS@RT, and Ce_SP@RT) sonication products is discussed first. Based on the physical state of these products, a set of representative experimental data is collected and is presented below. These set of experimental data establishes the efficacy of ultrasonic cavitation induced product size reduction and dispersion devised synthetic protocol.

## Charge Transfer (CT) transition:

In this context, the Ce_Sl@RT product and its response to the introduction of another solvent $H_2O_2$ is followed by UV-Vis spectroscopy. The obtained spectral data sets are plotted and shown in the figs.3.1 (a)-(c). The recorded spectral transmission curves (STC) of the Ce_Sl@RT product; is a sigmoidal shaped curve. The data shape of the STC is a manifestation of the transition from 100 % UV-A range (10-400 nm) absorption to almost 100 % visible wavelength transmission, interconnected by a finite width charge-transfer (CT, $O^{2-} \rightarrow Ce^{4+}$) edge associated with the $O_{2p^6} \rightarrow Ce_{4f^0}$ electronic transition. The electrons gained by the cerium atoms, chemically is a reduction process. Thus, an increased content of reduced cerium atoms will bring anisotropy to the sigmoidal shape of the STC, leading to a red shift in the transmission edge, tailing deeper into the visible wavelength range. As a consequence, this type of nc-ceria material is a probable candidate for visible-light photocatalytic applications. This aspect is inherently contained for ceria in its variety of nanostructured morphological formulations.

The STC of a micron grain-sized ceria (coded as Ce_micron, blue-colored curve) considered to comprise the lowest fraction of reduced cerium atoms is included in the plot in fig.3.1 (a) as the standard reference. A surface charge state transition is brought by introducing equivalent volume fraction of $H_2O_2$ into the Ce_Sl@RT product. The transmission edge of this modified Ce_Sl@RT product STC (coded as SPW $H_2O_2$, red-colored curve) almost becomes coincident to that of the Ce_micron. This implies near-complete oxidation of the Ce_Sl@RT product. It should be noted that, for all of these UV-Vis spectral data acquisitions, the DI-water is the used solvent chosen to disperse and is considered as the baseline run.

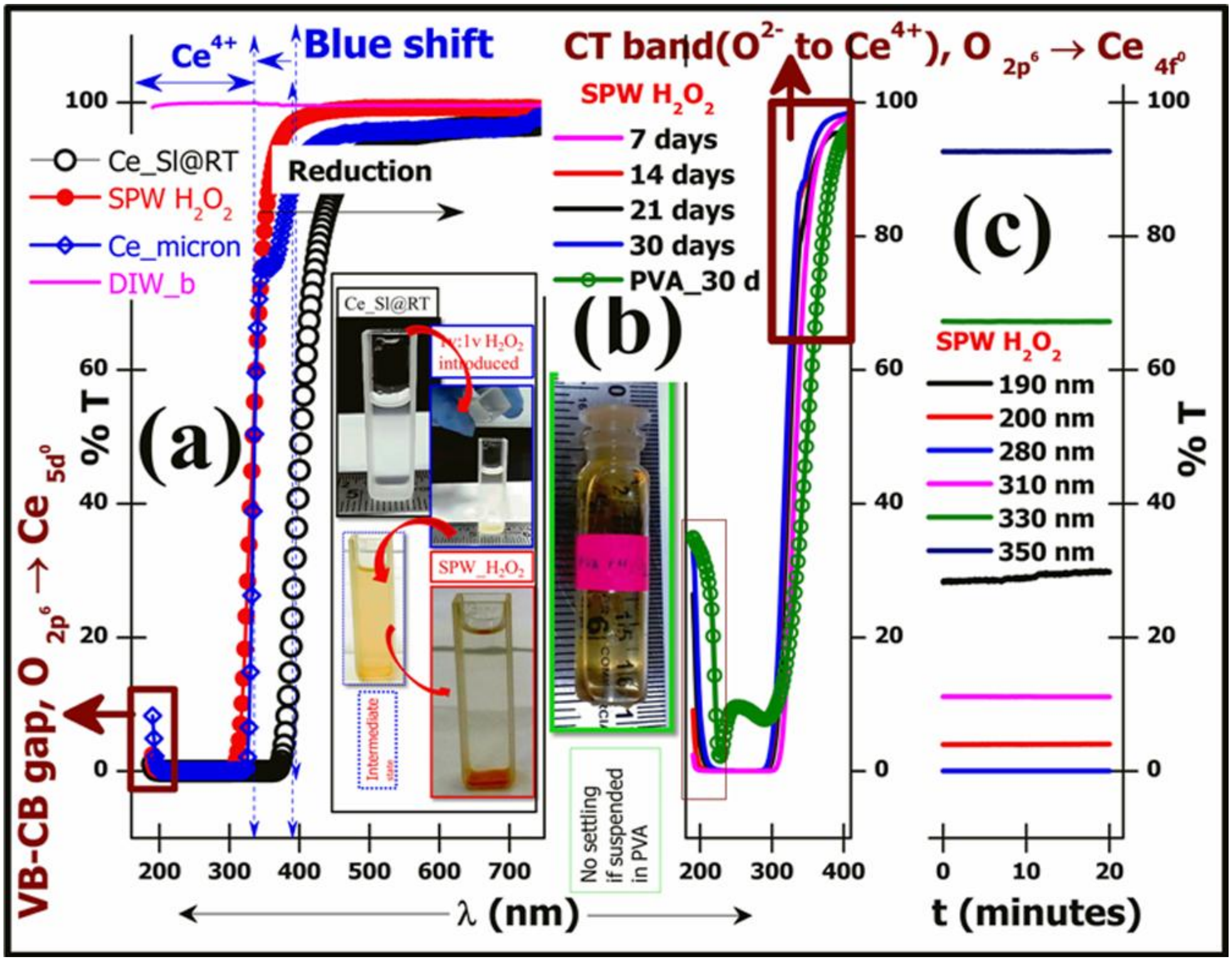

Fig.3.1 [***UV-Vis optical spectra of the Ce_Sl@RT***]: (a) The spectral transmission curves for original (Ce_Sl@RT) and interface modified (SPW $H_2O_2$) sample (b) SPW $H_2O_2$ aging stability, and (c) time dependent wavelength specific SPW $H_2O_2$ product measurements respectively.

The DI-water baseline trace (coded as DIW_b, maroon-colored) for the wavelength range analyzed has 100 % T and is also included in the plot. A set of transmittance kinks in the wavelength range=190-210 nm is observed and is assigned to the valency to the conduction band (VB to CB) electronic transition of ceria. The inset of fig.3.1 (a) depicts cuvettes containing solution dispersions in their chemical states before, during, and after oxidation in sequence, respectively. A fractional mass of nc-ceria that settles at the bottom of the cuvette is isolated, with aging in the $H_2O_2$ modified dispersion. After oxidation, the STC of obtained supernatant (coded as SPW $H_2O_2$) with days of aging is also recorded and is plotted in the figs.3.1 (b)-(c), respectively. Inset of fig.3.1 (b) is included to demonstrate that the Ce_Sl@RT has oxidized inside the polyvinyl alcohol (PVA) matrix. The oxidation occurs inside the PVA matrix, due to which settling derived mass is observed. The oxidation induced bright yellowish transparent coloration seen is retained for more than a year [57–62].

The optical band gap values and correlation with the changes in absorption edge brought forward by diluting the Ce_Sl@RT with equivolume fraction of $H_2O_2$ is discussed now. This is significant because both the Ce_Sl@RT and SPW $H_2O_2$ products are of the same particle size dispersions (PSD), but one is in reduced state while the other has surface cerium atoms in the oxidized state. This fact can be conclusively inferred by carrying out the oxidation process inside the polyvinyl alcohol (PVA) matrix. With precisely the same PSD, the surface oxidation achieved inside the PVA matrix (Fig.3.1 (b), the PVA_30 d sample has almost the STC as that of the SPW $H_2O_2$). No particle settling and maintaining almost the same PSD is of significance. That is to say, $H_2O_2$ modification plays an important role in the surface oxidation of the cerium atoms, as is evident from the distinct STC of SPW $H_2O_2$ than Ce_Sl@RT. That is in ceria, the charged state of the surface cerium atoms will decide the absorption edge tailing tunably. The STC of Ce_micron is chosen as the standard representing the fully $Ce^{4+}$-oxidized state at ambient laboratory environments. In the present context, Ce_micron data will be viewed as calibration for oxidation and reduction as well as the standard reference pattern.

The STC data processing for the optical band gap is carried out based on literature [63–68]. In brief, the absorption coefficient ($\alpha$ $cm^{-1}$) is extracted from the STC and its derivative with respect to the energy exhibits a point of inflection in the absorption edge distinguishing the high Tauc and low Urbach- energy regions, respectively. The highest achievable adj. R-square close to 1 iterative fit to these above mentioned regions is carried out. The adopted mathematical expression; (1) Non-linear Belehradek for bandgap, and (2) linear fit for Urbach energy is used. The computed parameters (like bandgap, type of transition, point of inflection, band-tail width, and electron-phonon interaction strength) are listed in the table-3.1. Also, the representative Tauc plots for optical band gap and PSD for Ce_micron, Ce_Sl@RT, and SPW $H_2O_2$ are presented in figs.3.2 (a)-(b) respectively.

The main conclusions drawn from this analysis are presented now. The important outcomes are: (1) the blue shift in the STC in UV-A range is probably due to the CT optical transition associated with the valency fluctuations in surface oxidation states from $Ce^{3+}$↔$Ce^{4+}$. (2) It is evident from the coincident STC that the CT feature is not purely a surface effect as it is observed in both Ce_micron and SPW $H_2O_2$. The particle size distribution (PSD) data of the both (Ce_micron D50=1092 nm, SPW $H_2O_2$ D50=7.5 nm) is presented in fig.3.2 (b). (3) Other set of STC presented suggest the CT feature is related to the surface charge states than size.

**Table-3.1**: The optical parameters extracted from the measured STC for different samples

| Sample code | Point of inflection $[d\alpha/d(h\nu)=0]$ line width (lw) | Type of transition, Belehradek $Y=a(X-b)^c$ c=0.5, direct c=2, indirect | Band gap (eV) | PSD Poly dispersive Index (PDI) | Band-tail width & Electron-phonon interaction strength ($E_{e-ph}$) |
|---|---|---|---|---|---|
| **Ce_micron** | 3.85 eV<br>lw=0.2 eV | **c=0.5**<br>**R2=0.99769**<br>c=2<br>$R^2$=0.99867 | **3.16(1)**<br>0.70(4) | D10=718 nm<br>**D50=1092 nm**<br>**PDI=53 %**<br>D90=2224 nm | 66 meV<br>&<br>**Ee-ph=1.71** |
| **Ce_Sl@RT** | 3.35 eV<br>lw=0.1 eV | **c=0.5**<br>**$R^2$=0.9967**<br>c=2<br>R2=0.9813 | **3.30**<br>3.01(2) | D10=8.16 nm<br>**D50=13.6 nm**<br>**PDI=23.6 %**<br>D90=17.24 nm | 130 meV<br>&<br>**Ee-ph=3.33** |
| **SPW $H_2O_2$** | 4.05 eV<br>lw=0.1 eV | **c=0.5**<br>**$R^2$=0.9994**<br>c=2<br>$R^2$=0.9907 | **3.96**<br>3.58(3) | D10=5.58 nm<br>**D50=7.50 nm**<br>**PDI=30.8 %**<br>D90=10.5 nm | 170 meV<br>&<br>**Ee-ph=4.44** |
| **PVA_30 d**<br>Same as<br>**SPW $H_2O_2$**<br>Oxidation<br>inside PVA | 3.76 eV<br>lw=0.3 eV | **c=0.5**<br>**$R^2$=0.9991**<br>c=2<br>$R^2$=0.9845 | **3.64**<br>2.73(4) | D10=8.14 nm<br>**D50=18.5 nm**<br>**PDI=28 %**<br>D90=23.5 nm | 322.7(3.9) meV<br>&<br>**Ee-ph=8.37** |

The PSD data of Ce_Sl@RT (D50= 13.6 nm) is close to that of the PVA_30 d (D50= 18.5 nm). However, it needs to be further supported by other spectroscopic probes. (4) Lastly, the data analysis suggests that the CT optical absorption is due to a direct electronic transfer rather than an indirect one. The Levenberg–Marquardt algorithm iteration deduced Belehradek fit to identify the type of optical transition ($Y=a(X-b)^{c}$; whether c=0.5 (direct) or 2 (indirect)) is plotted for PVA_30 d to illustrate the direct optical transition as the most appropriate, and is shown below in the figs.3.3 (a)-(b). These current observations are in concurrence with those reported in literature [69–75].

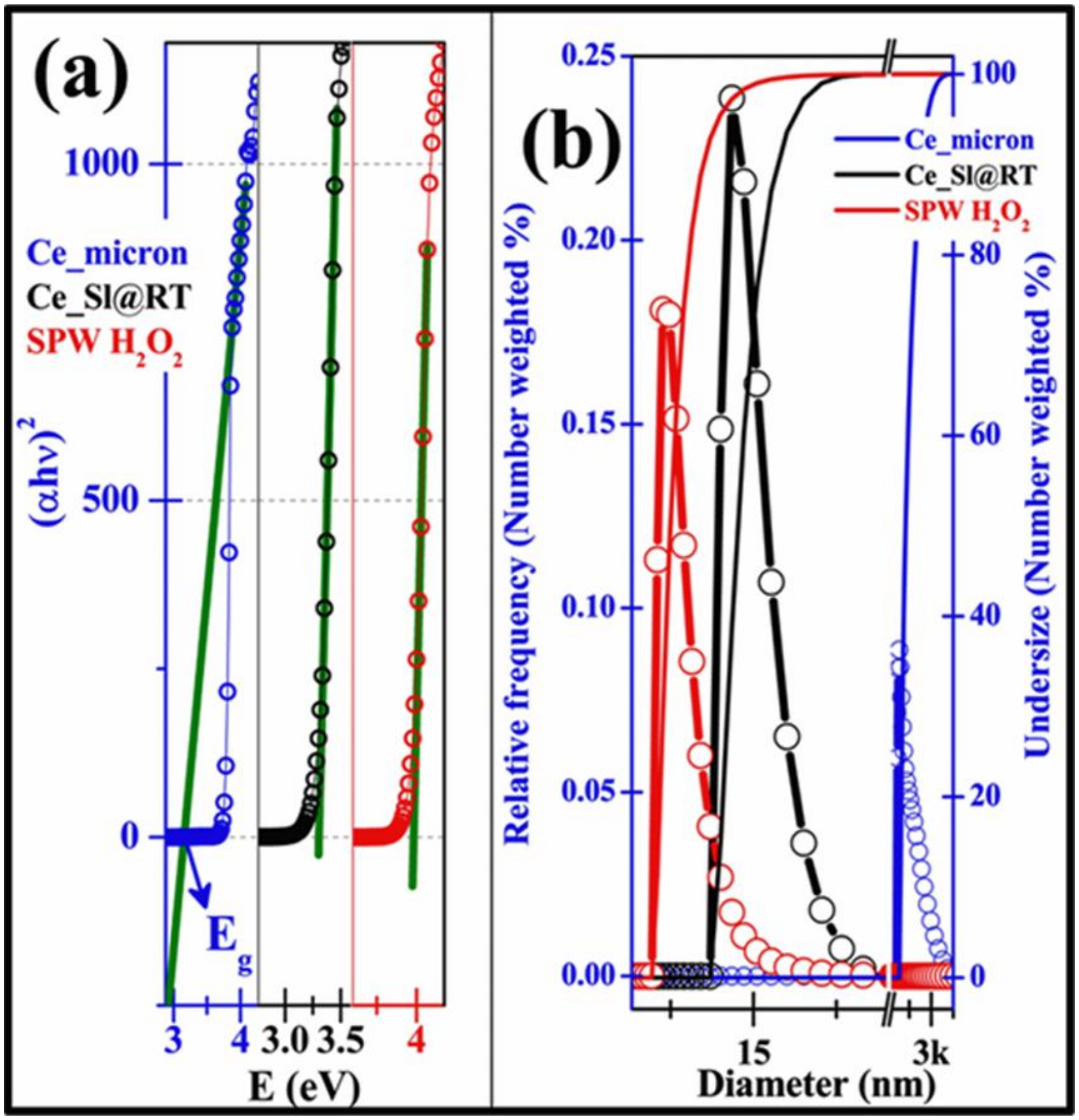

**Fig.3.2** [***Tauc plots and PSD***]: (a) Optical band gap plots, and (b) the corresponding PSD data respectively.

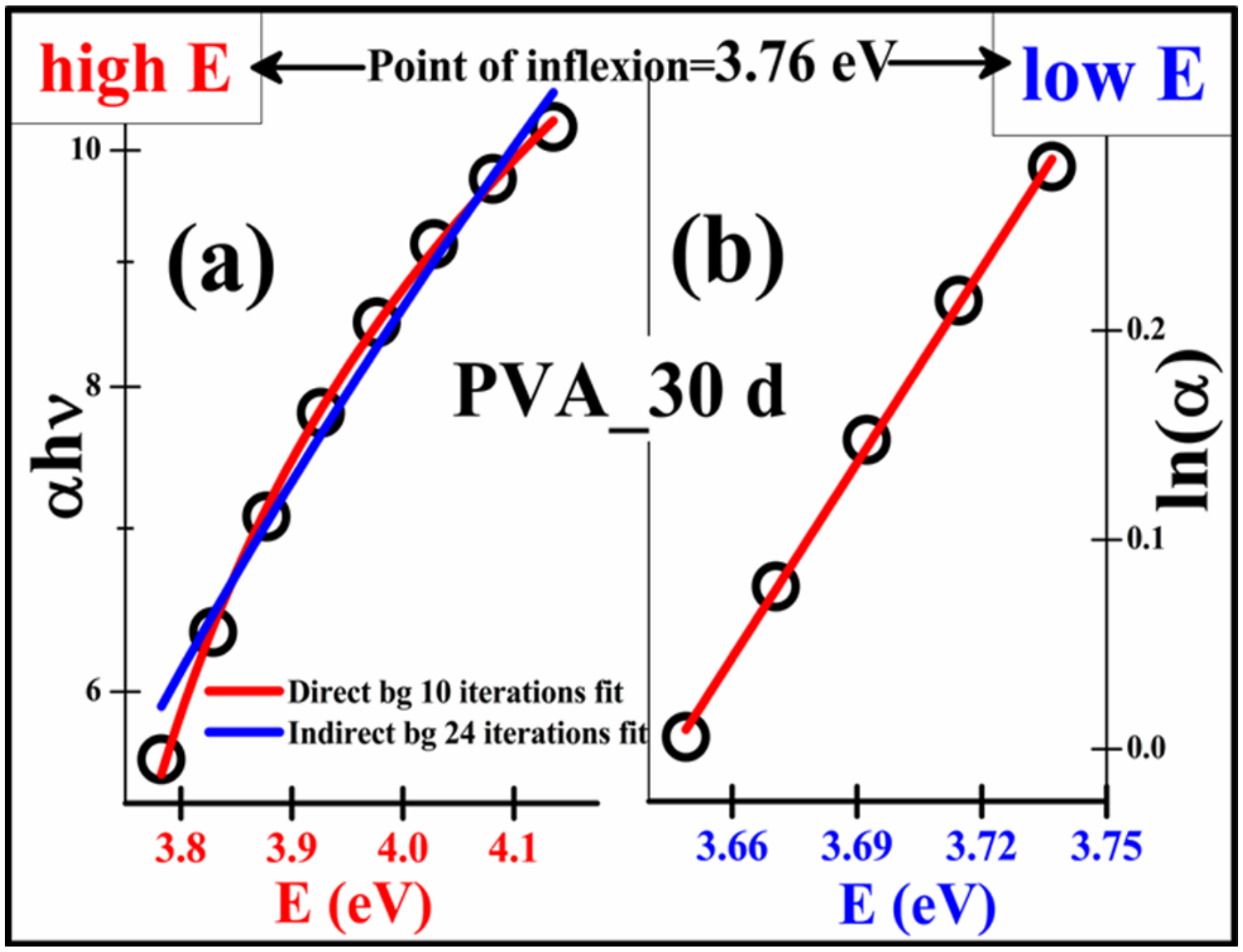

**Fig.3.3** [***Tauc and Urbach regions***]: (a) identification of the optical transition type, and (b) linear fit for band tail width and $E_{e\text{-}ph}$ strength computation respectively.

As stated in the previous paragraph that the origin of the observed blue shift, whether it is due to (1) surface CT feature, (2) quantum-size effect, or (3) a combination of both require further support by other subsidiary experimentation techniques which will be presented later. The gel-like Ce_SS@RT separated after supernatant liquid extraction during the process of the synthesis is used to address the above mentioned queries employing Raman spectroscopy. In the case of the Ce_SS@RT developed in the aqueous medium, the neighboring environment of a localized region can be easily changed by the addition of a few drops of $H_2O_2$ to bring about a CT transition of surface cerium atoms at RT with no effect all together on the PSD. The localized, oxidized and unoxidized regions with the Ce_SS@RT spread over an Al-foil wrapped BSG glass substrate can be probed employing the confocal-Raman spectroscopy system. The detailed analysis of such Raman spectral acquisitions in a single spectrum and time-series mode are discussed in the following sections.

## Origin of the blue-shift:

There has been much interest in the reversible switching feature in nc-ceria (crystallite size of about 7-8 nm or less, is the estimated Bohr exciton radius), in presence of different neighboring environments causing the autocatalytic regenerative CT transition of the surface cerium atoms (whether it is $Ce^{4+}$/or at $Ce^{3+}$) [76–81]. Most of the physicochemical and technological applications of nc-ceria are derived out of its tunable cyclic-CT feature. Based on this, the potential industrial utilities evolved are in the field of chemical mechanical planarization and polishing, catalysis, solid-oxide fuel cells of intermediate-temperature, and also activities in the field of sensors respectively [82–86]. However, there are contradictory reports that the reduction of the crystallite size will lead to increase in $Ce^{3+}$ concentration (27 % enhancement observed from size= 30 to 3 nm change) as other literature suggest that the same 3 nm nc-ceria crystallites don't have any $Ce^{3+}$ [87,88]. It underlines the importance of processing protocol in delivering either the fully oxidized or reduced cerium surface atoms in Bohr-excitonic sized nc-ceria formulations.

To achieve greater insight into the CT process, Ce_SS@RT is painted on a piece of gray emery paper to illustrate the adaptability to the changes in neighbouring environment. The painted Ce_SS@RT is colorless as seen in fig.3.4 (a). The color change of this gel is adjudged with respect to white-colored Ce_micron powder, which is in the $Ce^{4+}$ state based on the STC observations of the fig.3.1 (a). The addition of few DI-water drops to the painted Ce_SS@RT region changes the gel color to white, indicating that DI-water acts as oxidant. It is particularly noteworthy to mention that in the earlier discussion, the separated supernatant DI-water nc-ceria dispersion

(Ce_Sl@RT) ceria crystallites are in a reduced state. This, thus, confirms large number of investigations that state "nc-ceria oxidation state is transient and regenerative in nature predominantly decided by the adjoining and neighboring environment changes." Due to this transient nature of the cerium oxidation state in nc-ceria composite formulations, an important requirement is to specify by experimental means the cerium oxidation state before any technological use. Below presented experiment demonstrates this characteristic oxidation-reduction transition. A cycle of oxidation and its recovery, also termed as the auto regenerative feature of the nc-ceria at ambient laboratory condition with time evolved snapshots, are shown in the figs.3.4 (a)-(d). Out of many possibilities, the probable surface cerium atomic charge states most suitable for the present scenario is also schematically depicted for easy understanding, as adapted from literature [89–92].

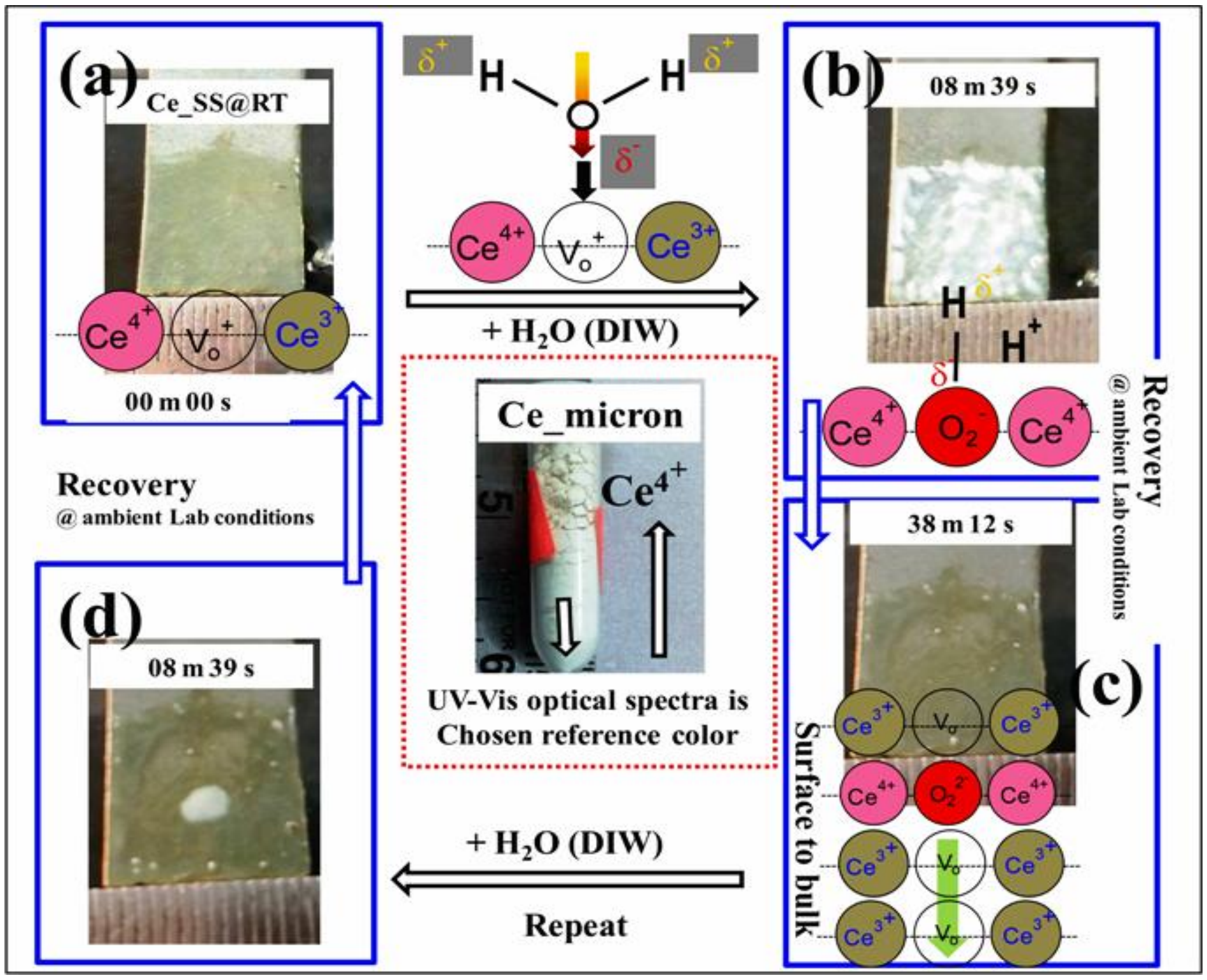

**Fig.3.4** [*Autocatalytic regenerative feature*]: The time-evolved oxidative and reductive feature of Ce_SS@RT, snapshots in DI-water at ambient laboratory conditions (inset Ce_micron powder in an eppendorf tube, surface schematics are adapted from reference [89–92]).

In this section the interpretation of Raman spectral signatures of the Ce_SS@RT and that of the dried solid powder (namely Ce_SP@RT) obtained in the CT transition context is discussed. The Ce_SP@RT Raman data as a function of the excitation wavelength ($\lambda_{excitation}$=532, 633, and 785 nm are used) is presented in fig.3.5 (a), which clearly reveals the 464 $cm^{-1}$ (wagging vibration of the oxygen atom between two $Ce^{4+}$ ions) a characteristic Raman signature of the ceria [93–98]. For the case of the Ce_SP@RT, the observed broad significant photoluminescence (PL) band at 636 nm is of interest. The fig.3.5 (c) demonstrates the dominant nature of 636 nm PL band on the 545 nm Raman spectral signature as obtained with the 532 nm excitation. The PL suppression is achieved with 785 nm excitation, i.e., moving away from the band gap or sub-band gap absorption observed by the 532 nm laser excitation. The physical illustration of PL suppression achieved using 3 different excitation sources is presented in fig.3.5 (a). Even for the 532 nm excitation, when the irradiance delivered to the Ce_SP@RT sample surface is 40 $kW/cm^2$ or above PL quenching is observed. In terms of power value when the localized region is exposed at 2.5 mW for continuously 12 minutes, the 545 nm Raman signature overtakes the PL signal. This dynamic changes highlighting PL quenching at the laser exposed region is continuously acquired in time series mode and is plotted in fig.3.5 (c). The Ce_SP@RT surface in pristine, with laser ON spot size, and after 12 minutes laser modified regions by 532 nm exposure are presented in fig.3.5 (b). Appearance of bright yellowish coloration compared to pristine sample with exposure hints at the possible surface modification. If the surface modification happens to be recoverable, the PL must recover which will be presented in later discussions. Thus the Raman spectroscopy conclusions for the Ce_SP@RT are; (A) A broad intense 636 nm emission overtakes small 545 nm Raman signature, (B) the 636 nm PL is quenched by two ways-(1) by moving away from the band gap or the sub-band gap absorption observed for 532 nm, (2) by exploiting the surface modification using laser output power value on the sample. The experimental probing on intervalence CT absorption in cerium (III, IV) oxide system having a broad, intense absorption band centered around 16300 $cm^{-1}$ (~ 613 nm) as reported by Allen et.al. [99]. In this work it was illustrated that the cerium (III) oxide absorption band at 613 nm undergoes quenching by air oxidation at RT. Also, the various other possible CT transition bands associated with cerium are listed in the specific references in table-3.2.

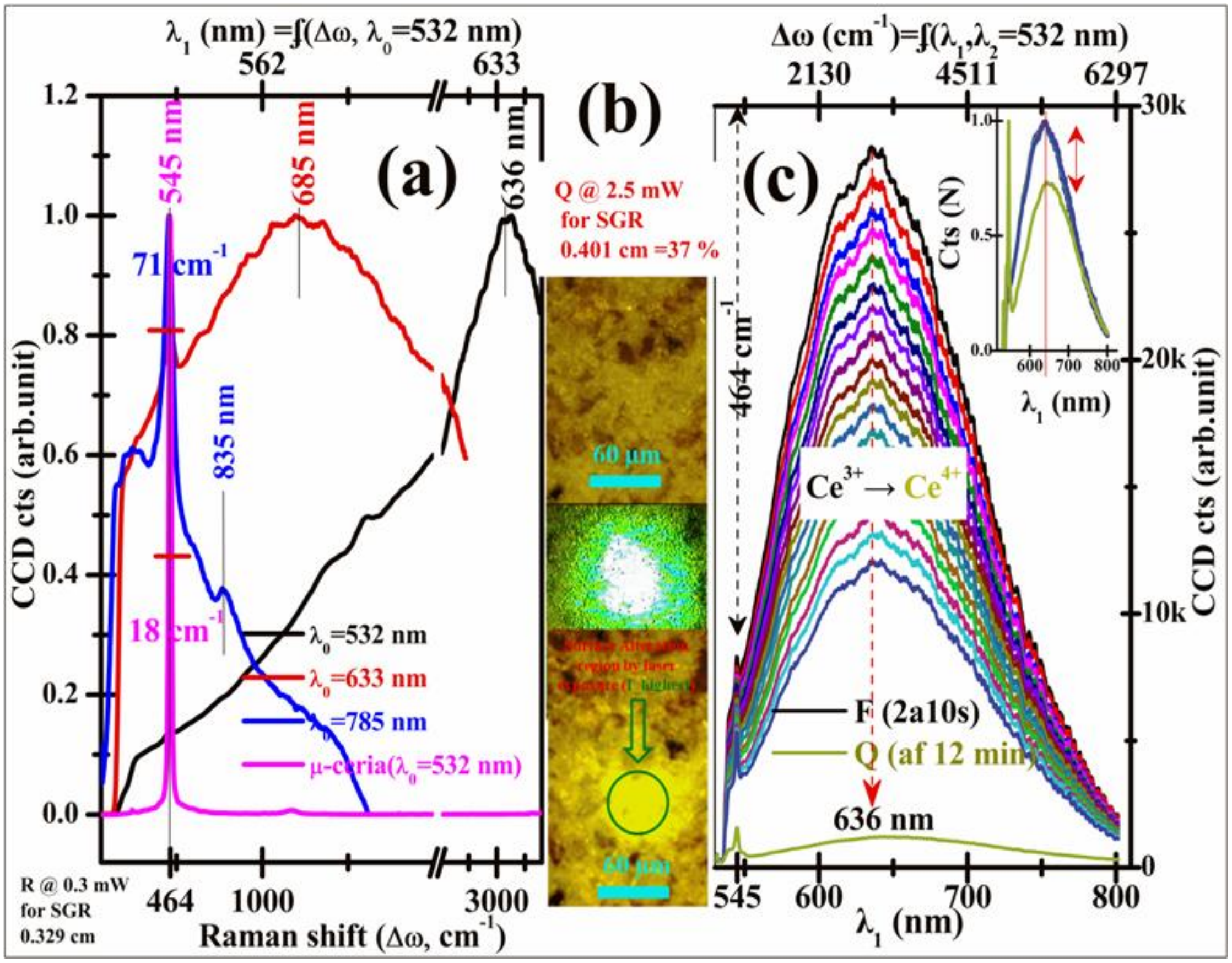

**Fig.3.5** [***Ce_SP@RT product CT luminescence***]: (a) Separating the elastic scattered Raman phonon signature from inelastic PL photons by different laser excitation wavelength, (b) Optical micrographs of Ce_SP@RT; $\lambda_o$=532 nm laser surface modified region of Ce_SP@RT product (A 20x 0.4 NA objective is used to focus the laser to about 3-5 μm spot), and (c) Quenching of the CT luminescence band employing $\lambda_o$=532 nm laser operated at 2.5 mW output maintained for 12 minutes demonstrating nc-ceria surface oxidation respectively.

**Table-3.2:** CT absorption band position in nm for the case of nc-ceria (reported and this work).

| Nature of species and transition (nm) | | | | | References |
| --- | --- | --- | --- | --- | --- |
| $Ce^{3+}$ & $f \rightarrow d$ | S: $Ce^{3+}$-$O^{2-}$ & $O^{2-} \rightarrow Ce^{3+}$ | S: $Ce^{4+}$-$O^{2-}$ & $O^{2-} \rightarrow Ce^{4+}$ | Interband transition | S: $Ce^{3+}/Ce^{4+}$ & $Ce^{3+} \rightarrow Ce^{4+}$ | S for surface |
| 200-250 | 263 | 280 | 320-350 | 588 | [100] |
| 208-218 | 250 | 297 | 320-340 | ------ | [101] |
| | | | | | |
| **233** | **------** | **299** | **------** | **407, 485** | **Ce@RT** |
| | **$E_{gd}$=1.56 eV** | | | **$E_{gd}$=3.77 eV** | **This Work** |
| **------** | **------** | **------** | **358** | **426,588** | **Ce_SP@RT** |
| | **$E_{gd}$=1.32 eV** | | | **$E_{gd}$=2.34 eV** | **This Work** |

These initial inferences about the cerium (III, IV) oxide system absorption bands will assist in analyzing the present case of Ce@RT and Ce_SP@RT products. The recorded optical absorption spectra of these products are plotted in fig.3.6 (a). The data is fitted to multiple nonlinear Gaussian peaks combinations to generate the best fit (adj. R-square close to 1). While fitting, the peak centers are chosen closest to the reported absorption peaks. However, peak width is allowed to vary without much overlap. After suitable fitting to the original data, the number of extracted absorption peaks specific to the samples was identified.

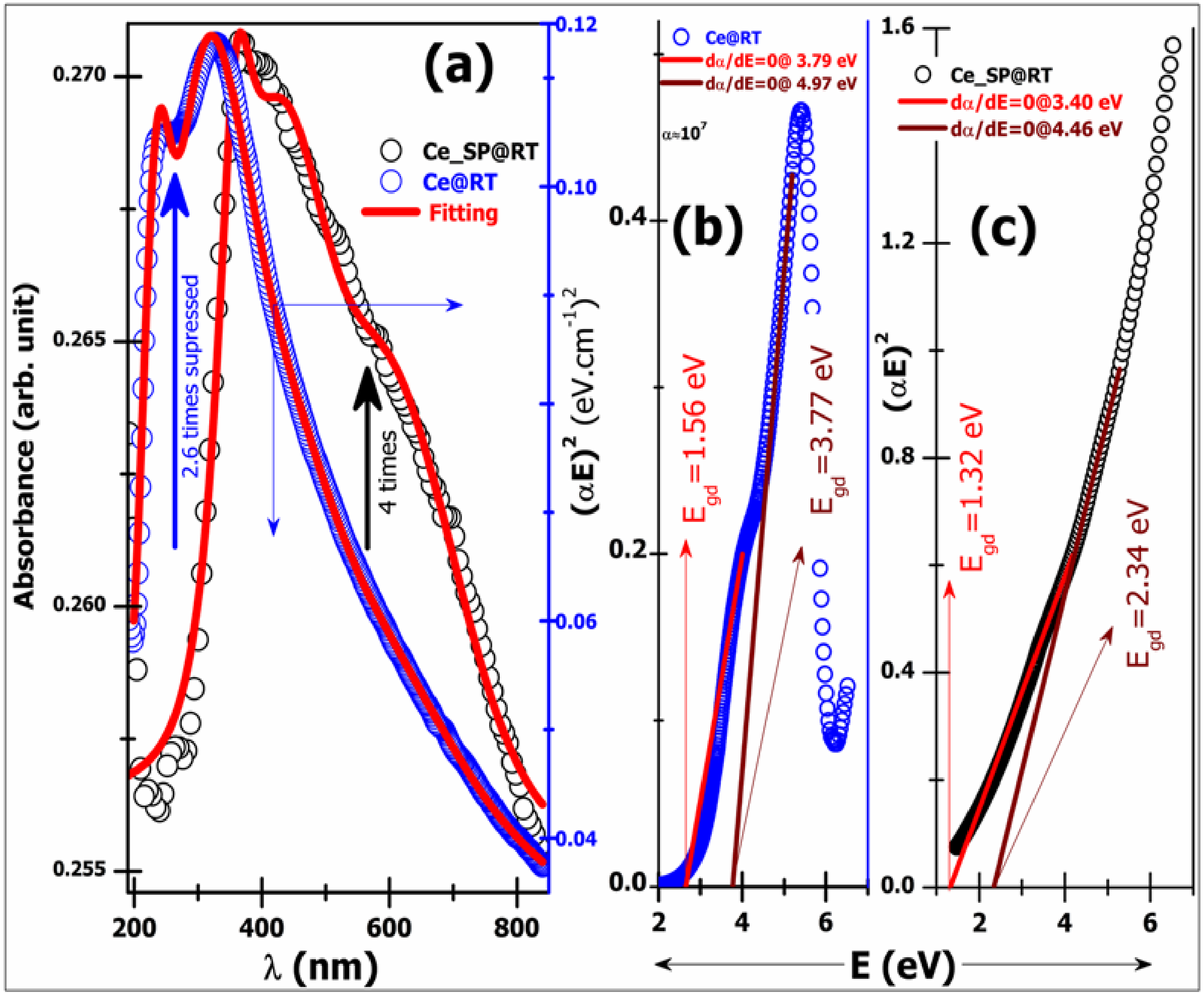

**Fig.3.6** [***Optical band gap Narrowed defective nc-ceria***]: (a) Optical absorption spectra of ceria products i.e., Ce_SP@RT and Ce@RT, (b) and (c) are the respective Tauc plots demonstrating computed bandgap values.

An obvious noticeable feature in the Ce_SP@RT product is the hike in absorbance efficiency, i.e., the low energy absorption tail ramps up to approximately ten times higher value (see fig.3.6 (a)) than the parent Ce@RT product. In contrast Ce@RT maintains its UV-A optical absorbance as evidenced by the blue dotted data curve of fig.3.6 (a). Literature suggest the absorption edge of reduced ceria must have a strongest band around 400 nm attributed to absorption by $Ce^{3+}$ at $C_2$ site. This is assisted by sidebands near 575 nm (absorption by $Ce^{3+}$ at $C_{3i}$ site) and 510 nm (absorption by $Ce^{3+}$ at $C_{3i}$ site) respectively [102]. The $C_{3i}$ sites oxidize more slowly than the $C_2$ sites. Likewise, for oxidized ceria products having $Ce^{4+}$ as a dominant fraction must exhibit an intense absorption onset below 370 nm [103]. The representative computed sub-optical band gap values (plotted in fig.3.6 (c)) reaffirms that the Ce_SP@RT product has defects. These defects fill the band gap with a variety of sub-levels that physically contribute to the optical band gap narrowing. One of the reasons for the observed redshift in the band gap is assigned to the increase of $Ce^{3+}$ fraction in the sample [104–107].

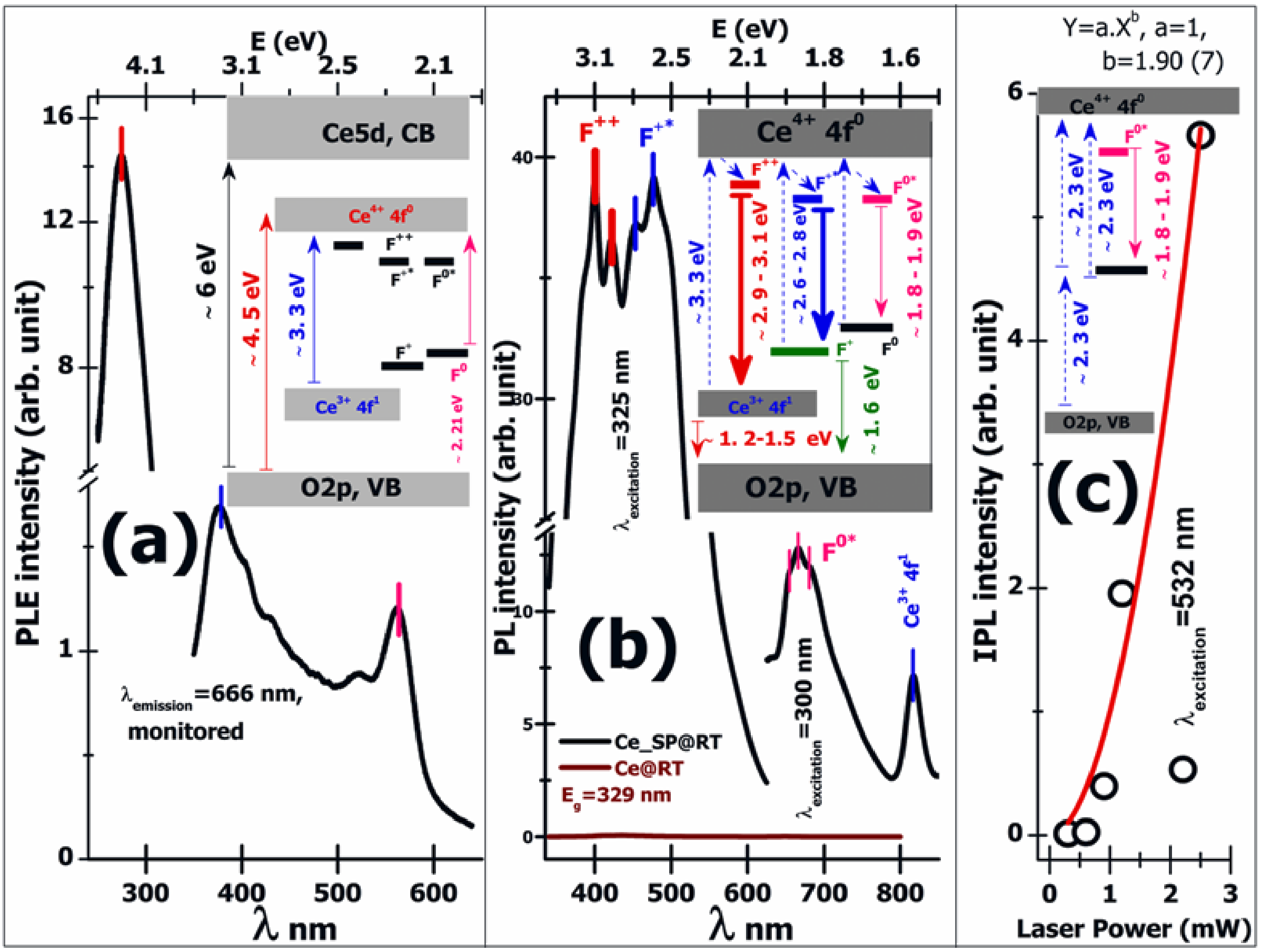

Fig.3.7 [*Photoluminescence spectroscopy of Ce_SP@RT*]: (a) PLE spectral data acquired for 666 nm PL (inset gives schematic of sub-optical gap absorption), (b) PL emission with probable defects, and (c) occurrence of the 2 photon absorption for 532 nm sub-band gap excitation respectively for Ce_SP@RT product.

The Ce_SP@RT product PL spectra in figs.3.7(a)-(c) justify the presence of sub-bandgap defect levels and assignments are consistent with literature [107–109]. The UV-VIS absorption spectroscopy in the earlier section the UV-Vis absorption spectroscopy confirms the sub-band gap absorption for Ce_SP@RT at or around 2.34 eV. The Raman spectral data presented in fig.5, acquired by 532 nm (2.33 eV) excitation with a broad intense emission at 636 nm is a direct affirmation. Further near band gap excitations implies a highly pronounced PL emission (fig.3.7 (b)) of Ce_SP@RT than that of quenched like PL emission data observed for the Ce@RT. The PL emission spectra recorded for both the Ce_SP@RT and Ce@RT products are for $\lambda_{excitations}$=325 and 300 nm, respectively) which is close to computed optical gap of 3.77 eV (329 nm) for Ce@RT. The intense major-violet and the blue emissions are $F^{++} \rightarrow Ce^{3+}4f^{1}$ and $F^{+*} \rightarrow F^{+}$ transitions, whereas the weak minor-red and IR emission is ascribed to the $F^{0*} \rightarrow F^{0}$ and $Ce^{3+}4f^{1} \rightarrow O_{2p}$, VB electronic relaxations respectively. The $Ce^{3+}4f^{1}$ acts as a hole trap and creates an energy gap of ~3.1 to 3.3 eV with an empty $Ce^{4+}4f^{0}$ sub-band. The symbolic representation of oxygen vacancy with two trapped electrons is $F^{0}$ while subsequent $F^{+}$ and $F^{++}$ states represent the loss of one and two electrons respectively. The excited states are $F^{+*}$ and $F^{0*}$, respectively. It is important to point out that the electron localization within the $Ce^{3+}4f^{1}$ band is not fixed to a single energy. In fact, a spread of multiple defect states are realized all along the band gap based on lattice distortion and availability of oxygen vacancies near to the $Ce^{3+}$ site [107]. In fig.3.7 (a) is the photoluminescence excitation (PLE) spectra of the PL emission $\lambda_{emission}$=666 nm excited by 300 nm excitation. The PLE curve has three absorption bands at (1) 275 nm ($O_{2p} \rightarrow Ce^{4+}4f^{0}$), (2) 375 nm ($Ce^{3+}4f^{1} \rightarrow Ce^{4+}4f^{0}$), and (3) 561 nm respectively. These PLE absorption band positions agree with that of the previous UV-Vis optical absorption data presented in

table-3.2. Thereby it becomes another supportive data to the defects induced bandgap narrowing feature. Even though the Ce_SP@RT bandgap is in direct coincidence with the 532 nm photon for excitation, the variation of the laser power and its effect on the integrated PL intensity (IPL, fig.3.7 (c)) indicates the possibility of two-photon absorption. Here, laser power ramps the IPL value approximately in square term ($LP=a\times IPL^2$). Thus, ultrasonic activation of the Ce@RT carried out at RT to develop Ce_SP@RT product induces sets of sub-bandgap defects levels responsible for the observed bandgap narrowing and the broad 666 nm PL emission.

The question that needs to be addressed, whether the ultrasonic processing even at RT facilitate the Ce@RT size reduction? Powder X-ray diffraction (XRD) is employed to answer this question. The XRD data of Ce@RT and Ce_SP@RT are presented in fig.3.8 (a), and analyzed employing the fundamental parameter approach (FPA) line profile fitting for the laboratory X-ray diffractometers [110]. The crystal structure refined calculated profile ($Y_{cal}$), difference pattern ($Y_{obs}$ - $Y_{cal}$), and the standard ICCD PDF: 81-0792 file Bragg positions as vertical lines are shown in fig.3.8 (a). The refined FCC unit cell lattice parameter for the Ce_SP@RT (with GOF ($\chi^2$) =1.33) and Ce@RT (with GOF ($\chi^2$) =1.27) are: a=5.443(2), and 5.425(2) Å respectively. The lattice expansion observed with respect to the standard a=5.412 Å imply both these products are under tensile strain. The strain value of Ce_SP@RT is 2.25 times higher than that of the Ce@RT. While plotting (see fig.3.8 (a)); the respective average strain values are scaled to zero, so that both XRD patterns are aligned with the standard ICCD PDF: 81-0792 file. XRD pattern of Ce_micron (bottom column fig.3.8 (a)) is also plotted, to demonstrate peaks significant FWHM broadening characteristic to these nano size products. The crystallites size reduction from 2.95 (2) to 1.96 (2) nm is observed. It supports the role of 20 kHz ultrasonic probe in size reduction [111]. Furthermore, no structural phase of oxygen non-stoichiometric is observed. The list of all non-stoichiometric ceria phases, and their stability at ambient/non-ambient conditions is tabulated by T. Alessandro et.al., [112]. Thus, the highly-strained fluorite unit-cell of the 2 nm crystallites representing Ce_SP@RT is in support with the optical band gap narrowing and PL sub-band gap excitation results.

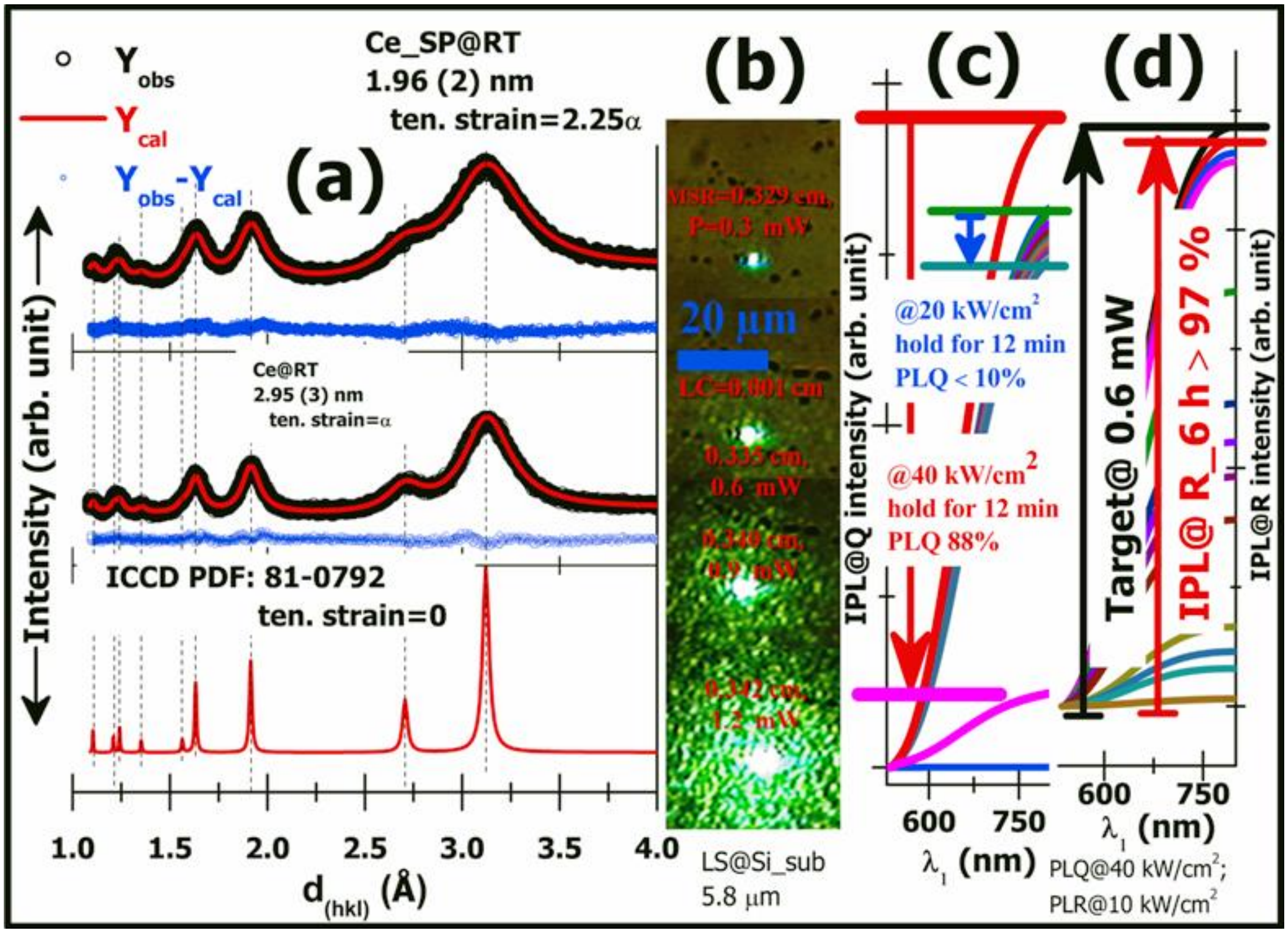

**Fig.3.8** [***Effect of the ultrasonication and 532 nm Nd-YAG laser irradiance on Ce_SP@RT product***]: (a) XRD data of ultrasonic induced product size reduction; (b), (c), and (d) are the IPL observed by laser irradiance of 5, 10, 15, 20, and 40 kW/cm$^2$ respectively. In increasing order these set of values represent PL-recovery, steady-state, and quenching data recordings.

The following conclusions can be drawn about the 532 nm Nd-YAG laser irradiance use to demonstrate the observed CT luminescence. The utilized laser power levels snapshots on the polished side of the silicon substrate are shown in fig.3.8 (b). The central laser spot area (neglecting the scattering generated spot spreading) is used for irradiance calculation. These are (1) 40 kW/cm$^2$ or higher irradiance is better for PL quenching, (2) between 5 to 20 kW/cm$^2$ is used for steady-state PL recordings (negligible change to IPL value in 535-800 nm range), and (3) 5-10 kW/cm$^2$ for IPL recovery. The plotted data in fig.3.8 (c) points to: Ce_SP@RT (a) IPL quenching and (b) for steady-state PL recording. Whereas (c) fig.3.8 (d) presents 12 minutes 40 kW/cm$^2$ irradiance modified region, subsequent IPL recovery data. IPL recovery of almost 97 % is observed after 6 hrs of hold in ambient conditions. These experimental illustrations confirm that for 2 nm Ce_SP@RT band-gap narrowed crystallites, the CT luminescence is a surface attribute. This IPL quenching and its recovery is ascribed to the neighbouring environment gaseous components either desorption/adsorption in a vice-versa way recurrently [113–121].

## Characterization of Ce_SS@RT product surface attributes using Raman spectroscopy:

The data presented in figs.3.4 (a)-(b), that demonstrated the autocatalytic regenerative feature of the Ce_SS@RT induced by the neighboring environment change (air to DI-water) is further explored by Raman spectroscopy. This study is expected to confirm, the introduction and depletion of DI-water as neighbour bringing the surface attribute changes of the Ce_SS@RT product. These data acquired at ambient condition is also useful to justify the earlier discussed CT transition. More specifically, the relative strength, presence, or absence of the Raman phonon modes over a micron-sized surface during the direct interaction with the oxidizing and reducing environment are in-situ mapped. This enables acquiring conclusive spectral and mapping information from the pristine and modified product surface region to illustrate the regenerative CT nature. The three specific Raman phonon modes of significance to the present nc-ceria surface analysis are first identified. These, with the increasing wavenumber are: (1) 460 cm$^{-1}$ (wagging vibration of oxygen atom between two $Ce^{4+}$ ions), (2) 560 cm$^{-1}$ (oxygen-stretching vibration between the $Ce^{3+}$ and $Ce^{4+}$ ions near to oxygen defects), and (3) 600 $cm^{-1}$ (stretching vibration of $M^{4+}$-O-$Ce^{4+}$ without oxygen defects), respectively [93–98]. The Raman mapping is carried out for 8 hrs at ambient condition, by which time the only changes to the nc-ceria surface occur while maintaining the PSD. The conclusion is that the observed phenomenon is solely a surface effect with no obvious dependence on particle size. Further, the physical significance of the above mentioned Raman modes in case of the metal-doped ceria ($M_xCe_{1-x}O_{2-d}$) composites, is extensively reported in literature and used to determine concentration of the (1) oxygen vacancies and (2) doped metal ions [93].

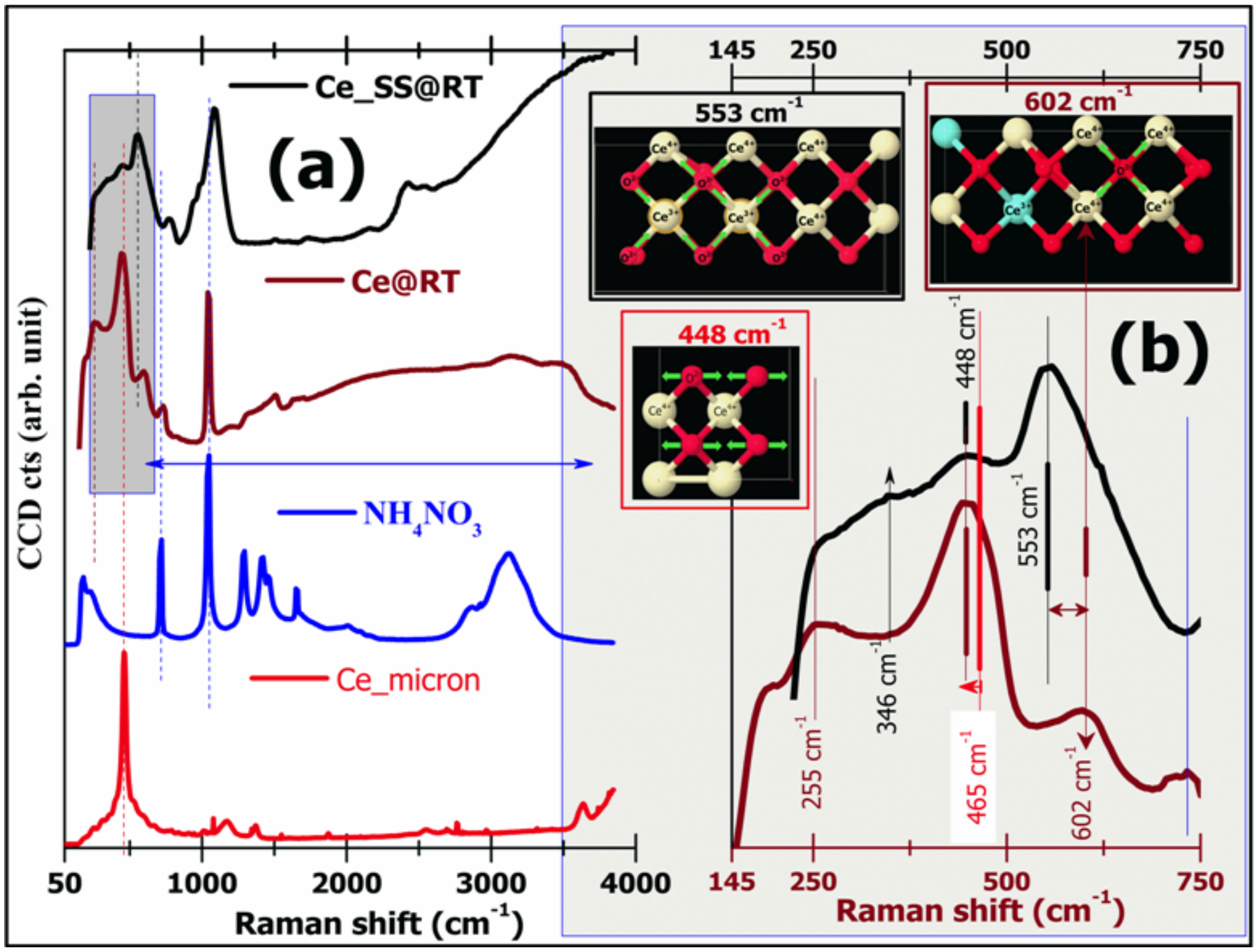

**Fig.3.9** [***Raman spectral phonon modes of the Ce_SS@RT product***]: (a) Ce_SS@RT Raman spectral features and its associated products, and (b) 448 $cm^{-1}$ (O-wagging), 553 $cm^{-1}$ (O-stretching near oxygen vacancy), and 602 $cm^{-1}$(O-stretching away from oxygen vacancy) phonon modes (schematic vibrations are adapted from reference [93]) respectively.

In the spectral data of fig.3.9 (a), observed most intense Ce_micron mode at 465 $cm^{-1}$ shift to 448 $cm^{-1}$ for Ce_SS@RT. It gets broader with low-energy asymmetric shoulder and is attributed to the combined effect of lattice phonon confinement and strain in nc-ceria [122–124]. The 488 $cm^{-1}$ mode in Ce_SS@RT and Ce@RT exhibits an inverse relationship with the corresponding oxygen vacancy related mode at 553 $cm^{-1}$ as adjudged from fig.3.9 (b) [i.e., as if appearance of one links to disappearance of other]. The reasoning reported is an introduced oxygen vacancy, in turn, leads to two-$Ce^{3+}$ ions (ionic radius of 1.143 Å) in place of two-$Ce^{4+}$ existing lattice ions (ionic radius of 0.97 Å). It in turn satisfies the charge neutrality, but brings an inbuilt tensile strain causing lattice expansion. The 2 nm Ce_SS@RT higher strained crystallites 488 $cm^{-1}$ mode is of much reduced intensity, and also have substantial peak broadening. A red shift of 17 $cm^{-1}$ and broadening aspect of 488 $cm^{-1}$ is presented in fig.3.9 (b). Likewise, the 602 $cm^{-1}$ mode with increasing concentration of oxygen vacancies red shift to 553 $cm^{-1}$ mode gradually and subsequently merges [122]. Besides these, a pronounced mode at 346 $cm^{-1}$ representing the sub-surface oxygen vacancies is seen, linked to the diminished surface O-Ce stretching vibration mode observed at 255 $cm^{-1}$. Other observed phonon modes at 782, and 1083 $cm^{-1}$ are also literature identified based on the neighboring environment and the synthesis conditions used [122]. The investigation of the nc-ceria structural phase transition by in-situ Raman spectroscopy is fascinating. It is to be pointed out that in literature the 276 $cm^{-1}$ phonon mode (representing along the c-axis both the Ce and O-atoms opposite movement) is monitored to address the tetragonality development. Thereby, a reversible tetragonal (P42/nmc) to cubic (Fm-3m) nc-ceria structural phase transition between -25 to -75 °C by monitoring its tunability was recently demonstrated [125]. The absence of the 276 $cm^{-1}$ mode in the present study confirms the XRD analysis that both materials are in cubic phase and no tetragonal phase is present. Thus, the current Raman spectroscopic investigation illustrates the inter-dependence of specific nc-ceria phonon modes, in which the appearance of the one directs the disappearance of the other and vice-versa. These dynamic in-situ phonon modes changes are mapped for the case of the Ce_SS@RT with respect to the pristine Ce@RT as standard.

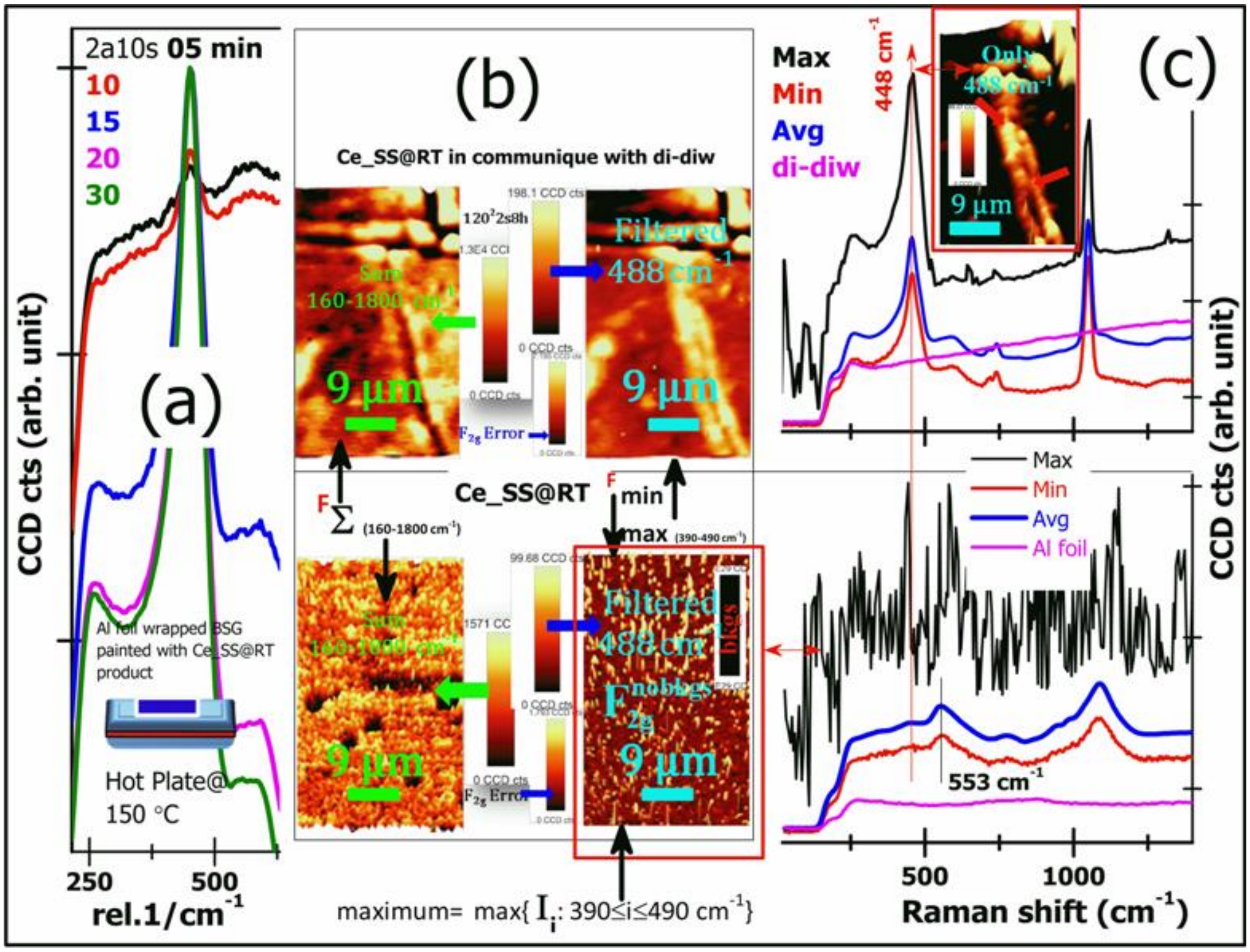

**Fig.3.10** [***Dynamic changes to the phonon-modes of Ce_SS@RT brought in by neighbouring environment changes***]: (a) thermal oxidation on hot plate @ 150 °C spectral traces, and (b) Raman mapping data in which the product is maintained in continuous contact with DI-water. (c) Spectral data extracted out of the mapping data to demonstrate the appearance of 488 $cm^{-1}$ mode respectively.

One of the standalone observations of the Ce_SS@RT product linked to Raman spectroscopy is highlighted here. The specific case of importance is the time-evolved oxygen vacancies elimination achieved in ambient by 150 °C hot plate annealing. In which the pristine broad diminished 488 $cm^{-1}$ mode recovers and strengthen with annealing time and saturates within 30 minutes. The probable reasoning for this observation might be linked to surface desorption of ambient environment species (like; hydroxyl, peroxides, superoxide's etc.) as literature reported leading to oxidation [113–121]. It confirms the possibility of localized laser-induced IPL thermal quenching, when investigated at 40 $kW/cm^2$ laser irradiance. The spectral data presented in fig.3.5 (c) at this irradiance level from the same product location demonstrates the CT luminescence quenching. Also, the in-situ Raman mapping data acquired from the 50 $\mu m^2$ Ce_SS@RT product surface during 8 hrs DI-water in contact will further reaffirm, whether the CT quenching is surface oxidation or not. The adopted nc-ceria surface schematic presented earlier ( see the inset of figs.3.4 (b)-(c)), highlights the recoverable surface oxidation scheme in contact with DI-water is published elsewhere [79,90–92].

An aluminum foil wrapped optically flat micro-glass (76×25×1.35 $mm^3$ borosilicate glass) slides painted with the Ce_SS@RT gel is chosen for the case of thermal healing in air by annealing. A set of 5 such Ce_SS@RT gel paintings are placed on a hot plate at 150 °C in the ambient condition and are withdrawn in intervals of 5 minutes individually. These samples are immediately taken for Raman acquisitions till a noticeable change of the 488 $cm^{-1}$ phonon mode is observed. The obtained individual Raman spectral data (2 accumulations each of 5 s) is normalized and are plotted in the fig.3.10 (a) for interpretation. A strong, intense 488 $cm^{-1}$ mode assigned to the nc-ceria oxidation started to appear with annealing time. Based on this observation, the observed IPL quenching by laser irradiance generated localized heating cannot be ruled out. Also, the irradiance localized region appearance changes in respect to the pristine region of the Ce_SS@RT product add favorably to surface desorption. The optical micrograph of pristine and 40 $kW/cm^2$ irradiated surface region is presented in fig.3.5 (b). In the surface modified region this irradiance is maintained for 12 minutes to deliver noticeable 88 % IPL quenching (see fig.3.8 (c)). Thus, this confirm surface desorption and localized oxidation induced by Nd-YAG laser-generated heating. The 97 % recovery data to the original reduced auto-catalytically by retaining the sample position in ambient for 6 hrs is shown in fig.3.8 (d). In the oxidized product the intensity of the 488 $cm^{-1}$ mode overtakes the 553 $cm^{-1}$ mode. In contrast, although the interchangeability between these two modes is not established in the present study, oxidation occurrence by bringing in contact with DI-water and co-dependence between the modes can be seen in the fig.3.10 (c).

The in-situ dynamic observations of the Raman spectral signatures in contact with the neighboring ambient environment changes are discussed here. The conclusions are; (A) in case of Ce_SS@RT product: (1) The maximum spectrum filtered in range 390-490 $cm^{-1}$ out of the sum filter (160-1800 $cm^{-1}$) happens to be noise, thereby no signature of the strongest 488 $cm^{-1}$ mode is observed over the mapped region of 50 $\mu m^2$ . (2) Its Raman map is featureless with uniform contrast. (3) The average and the minimum spectra retraced out of the total wavenumber mapped, exhibits the 553 $cm^{-1}$ oxygen vacancy mode, in conformity with the nc-ceria defects (i.e., at reduced state). Similarly, (B) for the case of Ce_SS@RT in continuous contact with the DI-water: (1) the Raman spectrum of the trench region of this product surface, sum filtered (160-1800 $cm^{-1}$), having DI-water in it is mapped. (2) The 488 $cm^{-1}$ is considered and is extracted out of this sum filter data. It can be seen as having differential contrast, implying the presence of significant oxidized cerium fraction on the mapped surface. (3) Away from the trench boundaries there is less surface oxidation. (4) An appropriate background subtraction from all over the mapped surface region is done to highlight the trench boundaries explicitly. It is shown as an inset in the fig.3.10 (c). (5) The presence of the 448 $cm^{-1}$ mode even in the minimum spectra derived out of the entire 50 $\mu m^2$ region, illustrates the oxidative feature of the DI-water. These Raman spectroscopic mapping illustrations, along with the shown extracted spectra, are in support of the autocatalytic regenerative feature snapshots of the Ce_SS@RT product presented earlier in figs.3.4 (a)-(d).

## Agglomeration of Ceria nanocrystals a boon or bane: a detailed evaluation in tune with the non-classical crystal growth route

Primary entities like atoms/ions/molecules their agglomerates and subsequent agglomerate aggregation are the generic evolution stages of developing molecular entities into a coarser particle. In-situ high-resolution Transmission Electron Microscopy (TEM) imaging is dominantly employed to elucidate and track these morphological evolutions linked with the crystal growth. Even the unexplored phenomenon of initial nucleation stage has also recently been observed to proceed via a two-step non-classical pathway [126–130]. The best-known fact about the classical crystal growth (stepwise layer-by-layer monomer adsorption derived growth) process was modeled as early as 1927 by Kossel et al., [131]. In this model, the solution-phase growth of a faceted crystal is presented. Without going much into the details of the classical growth, it can be stated that there are two main physical attributes: (1) overall free energy minimization drives nucleation, (2) in contrast overall surface energy minimization of the system directs crystal growth. This second physical parameter is the particle coarsening step and is also termed as Ostwald's ripening [132]. The prevalent classical crystal growth major stages are followed from the thermally annealed micron-Al powder FESEM micrographs, which are presented in the figs.3.11 (c)-(f). The micrograph of fig.3.11 (c) depicts a coarser μ-particle surface packed with nanoscale crystallites (inset in the micrograph) and also having smaller particles (identified in the same micrograph as 1, 2, and 3) as its surface attachments. The surface roughness of this μ-Al particle can be seen in the inset of the fig.3.11 (c). The figs.3.11 (d) and (e) represent micrographs of the $LN_2$ temperature quenched product made out of the sample shown in fig.3.11 (c) particles held at 700 °C for 2 hrs. In fig.3.11 (c) the previously surface attached smaller particles 1, 2, and 3 can be seen undergoing Ostwald ripening when in contact with the larger one.

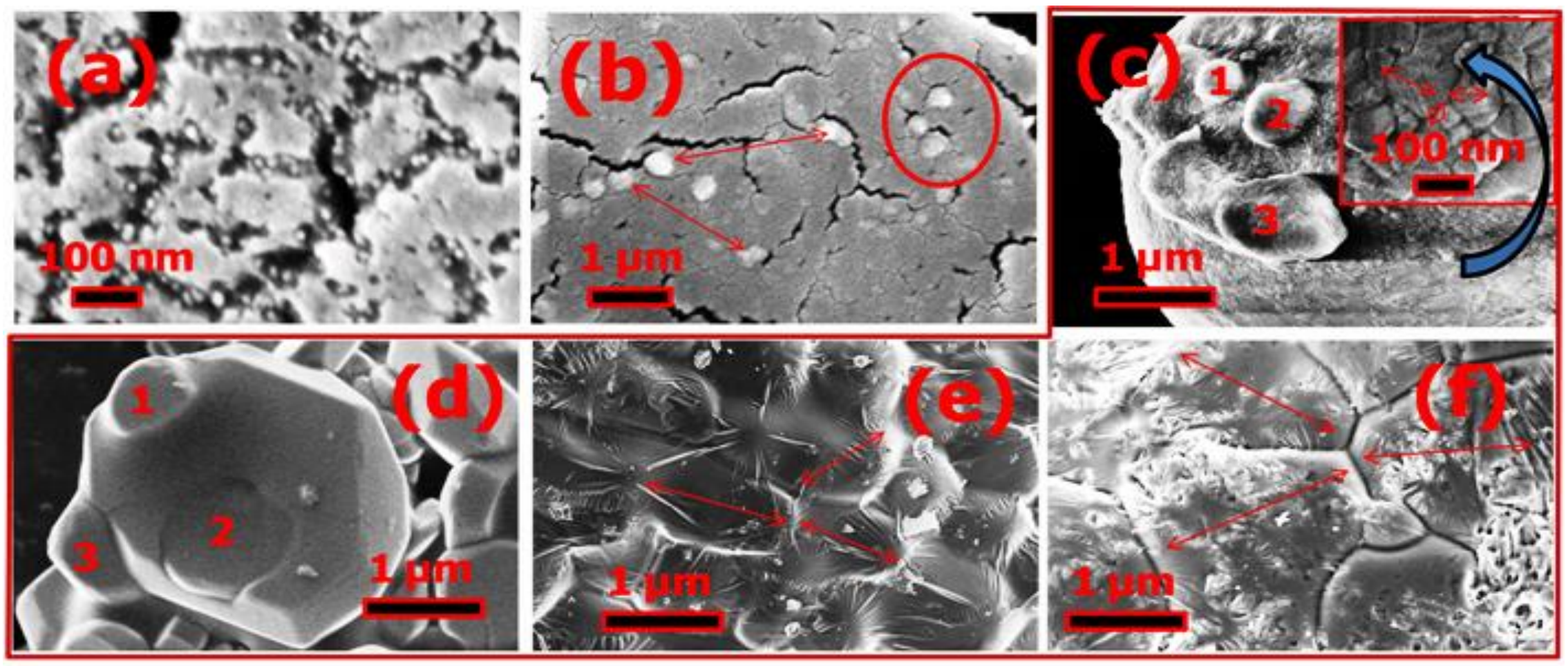

**Fig.3.11** [*Classical Crystal growth stages induced by thermal processing: FESEM illustration*]: Nucleation: (a) monomers generation and adsorption, and (b) nucleating sites evolution; Ostwald ripening: (c) micron-Al surface, (d)-(e) ripening, (f) final evolved grain surface with larger crystallites.

Likewise, the process of evolution of μ-Al particle surface with smaller crystallites to a much larger crystallite is depicted in fig.3.11 (e). Micrograph in fig.3.11 (f) is for the corresponding furnace cooled RT product of sample presented micrograph fig.3.11 (e). It highlights the actual surface evolution with a higher clarity. The micrographs in figs.3.11 (a)-(b) are acquired to present a schematic view of the nucleation stage with the following steps: (1) individual monomers adsorption with each other giving rise to an amorphous matrix, (2) a mass density fluctuation inside the amorphous matrix driven by the maintained external parameters will initiate some random nucleating centers as highlighted in the fig.3.11 (b). Nucleation is a single continual step (extremely small length $10^{-10}$ m and time $10^{-13}$ s scale). However it is divided into two micrographs for representative purposes to demonstrate it.

The micrographs from figs.3.12 (a)-(c) represent the individual building units: (a) 2D hexagon, (b) a cube, and (c) spherical particles. The morphologies observed in figs.3.12 (d)-(f) can be seen to be developed out of these smallest building units by systematic aggregation. In the available natural material today, about 90 % (specifically biominerals) belong to this depicted scheme of crystal growth. It is also termed as "the particle-mediated non-classical growth pathway". The subsequent discussions will be specific to non-classical crystal growth scheme of things leading to the primary question answer of whether nanocrystals agglomeration is a boon or bane. The most obvious response is that agglomeration is a boon, if it happens in a controlled fashion leading to fine-tuned targeted morphology evolution.

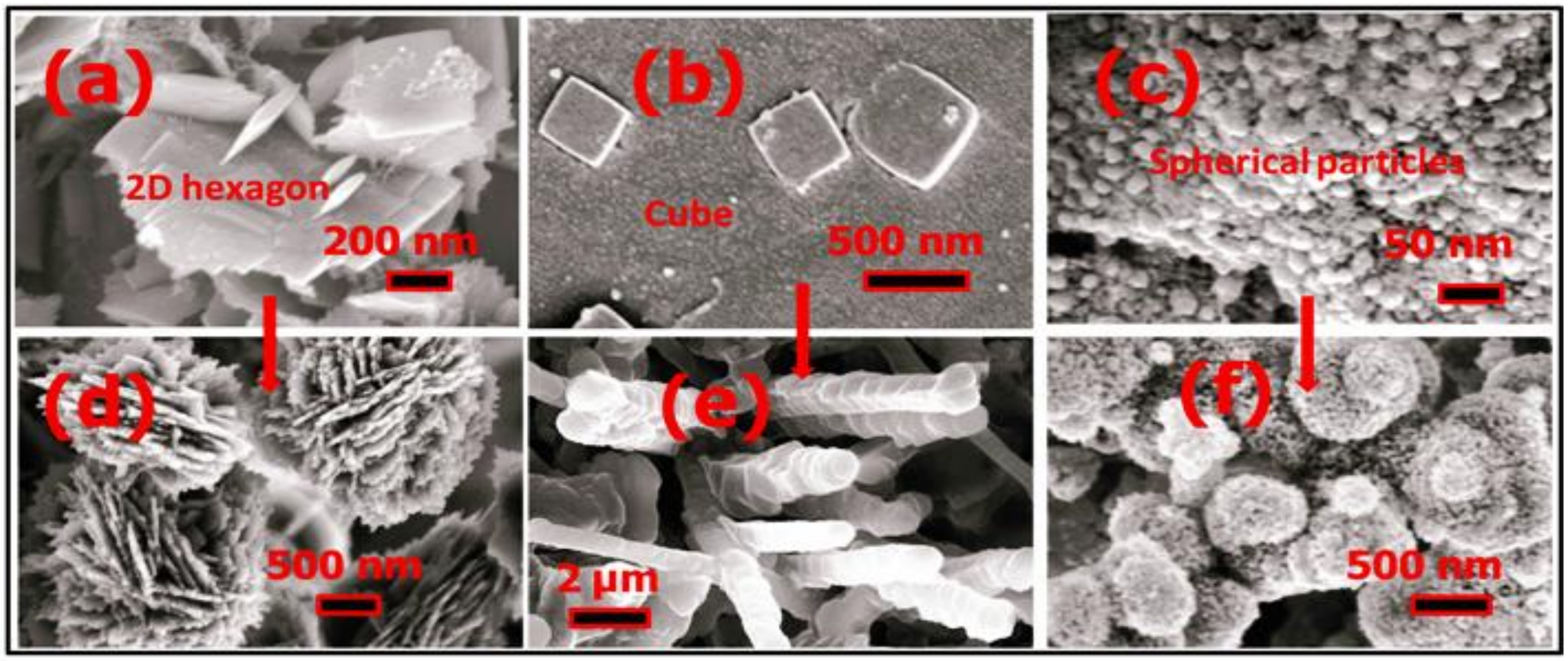

**Fig.3.12** [***Particle mediated two step non-classical crystal growth: FESEM illustration***]: Building blocks of: (a) 2D hexagon, (b) cubes, and (c) spherical particles; Evolved morphologies: (d) flower like, (e) rods, and (f) nano porous microspheres respectively.

In particle-mediated non-classical growth, observation of the current study and literature illustrates particles oriented attachment (OA). The OA between two building units or more usually advance by the fusion of the identical (hkl) highest surface energetic crystal facets. The energetic crystals facets are brought into contact with each other by undergoing appropriate rotation to develop into a single larger mesocrystalline feature. The detailing of such a mesocrystalline structures is mostly carried out by TEM-SAED mode investigation. However, in the present FESEM micrographs, the 2D hexagons will be mesocrystalline (see fig.3.12 (a)).

## 1D-Ceria fibers growth by OA mechanism in DI-water facilitated biomineralization:

As stated earlier, the freshly prepared Ce_Sl@RT product surface atomic cerium is in the +3 charge state. With aging for a year, the same supernatant Ce_Sl@RT auto recovers to its bulk +4 states gradually. The fresh supernatant and the aged supernatant will be denoted by Ce_Sl@RT (F) and Ce_Sl@RT (1Y), respectively, in the forthcoming discussions. The oxidation recovery of the surface cerium atoms with aging is confirmed by the blue-shift STC and can be seen in the fig.3.13 (a). This oxidation-reduction cycle is inherent to nc-ceria. Interestingly, settled entities at the bottom of the stored vial are isolated. These entities are DI-water facilitated OA driven bio-mineralized progressed 1D-ceria fibers. It is another experimental demonstration in support of the DI-water as the direct participant in mineralization, similar to many previous studies [24–26,28,118,133]. The fig.3.13 (b) inset shows the snapshots of the (1) 12 month settled mass (12 MSM), (2) its redispersion in the DI-water, and (3) EDS compositional map of these fibers drop cast on the carbon tape respectively. The sequential settled mass extracted after every 3 months' duration is investigated with the UV-Vis optical spectroscopy. The obtained UV-Vis spectral data of each of these products (i.e., 3 MSM, 6 MSM, 9 MSM, and 12 MSM) are plotted in the fig.3.13 (b). The absorption edge extension to the visible wavelength range with aging suggests the highly reduced state of the 12 MSM products. This is similar to the previously presented absorption data of the Ce_SP@RT.

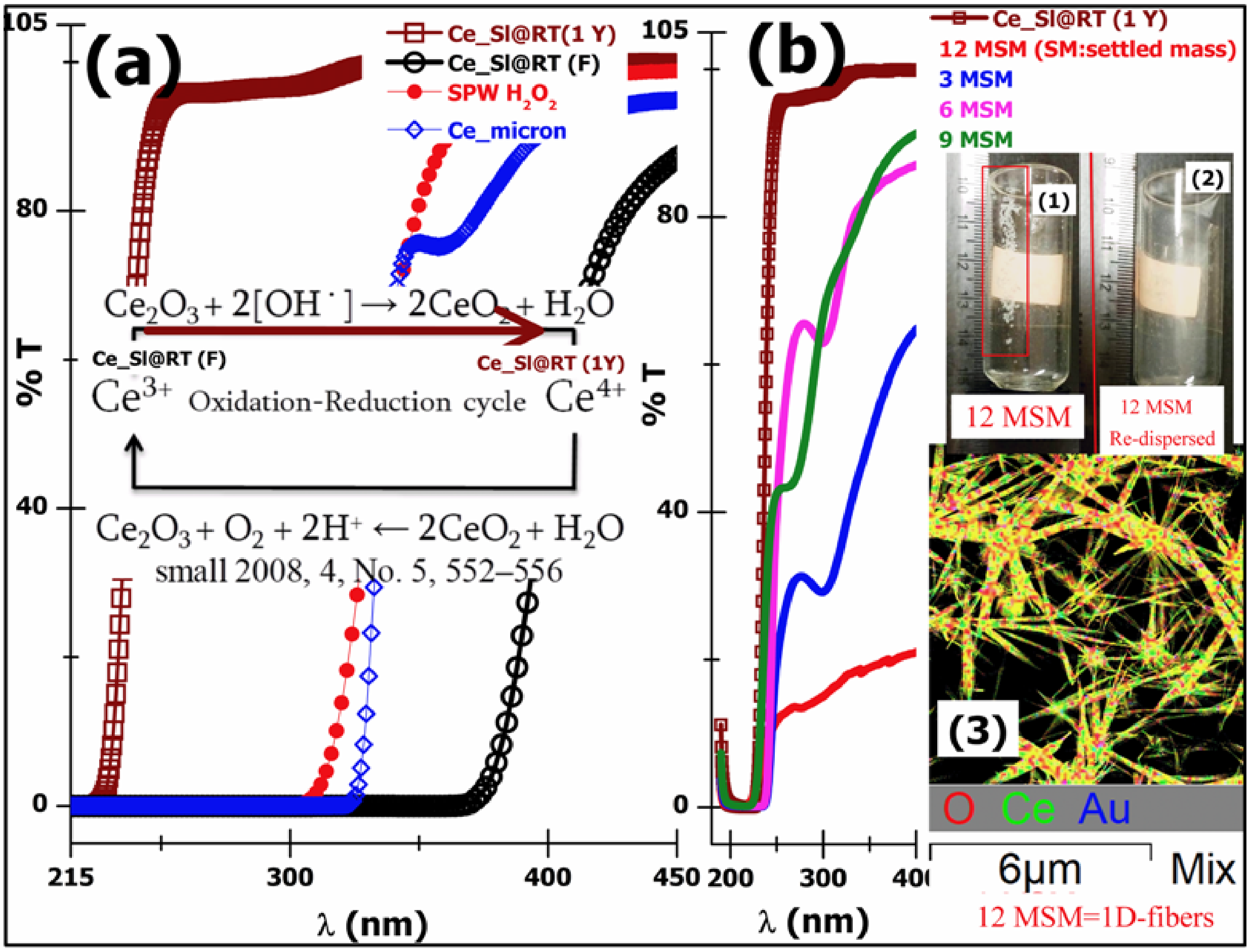

**Fig.3.13** [***Autocatalytic regenerative feature and biomineralization***]: (a) Oxidation by aging; (b) optical spectra of progressive bio-mineralized products (with inset of 12 MSM product in glass vial (1), redispersion (2), and (3) morphology with composition).

## Tyndall Effect:

The simplest way to illustrate the nc-ceria settling with aging is to visually track the laser light path intensity, which is allowed to travel through a set of nc-ceria dispersions. As can be seen, the 532 nm laser path is quite diminished when compared with the fresh, and the PVA stabilized nc-ceria dispersions. The PVA stabilized product had no scattering intensity loss, even after being stored for more than a year. It reasserts that even though the devised transparent supernatant has 2 nm ceria water-soluble crystallites, the DI-water facilitates these individual crystallites to interact in an orderly fashion and aggregate. The growth of these aggregates causes settling. The photograph of the Ce_Sl@RT product is presented in the figs.3.14 (a)-(b). It is evident from figs.3.14 (c)-(d) that the isolated 3 months settled mass (3 MSM) developed network type microstructure and its corresponding TEM micrograph concurrently favours DI-water as the growth directing agent. The direct participation of water in the biomineralization confirms reported earlier observations [11,24–26,28,133].

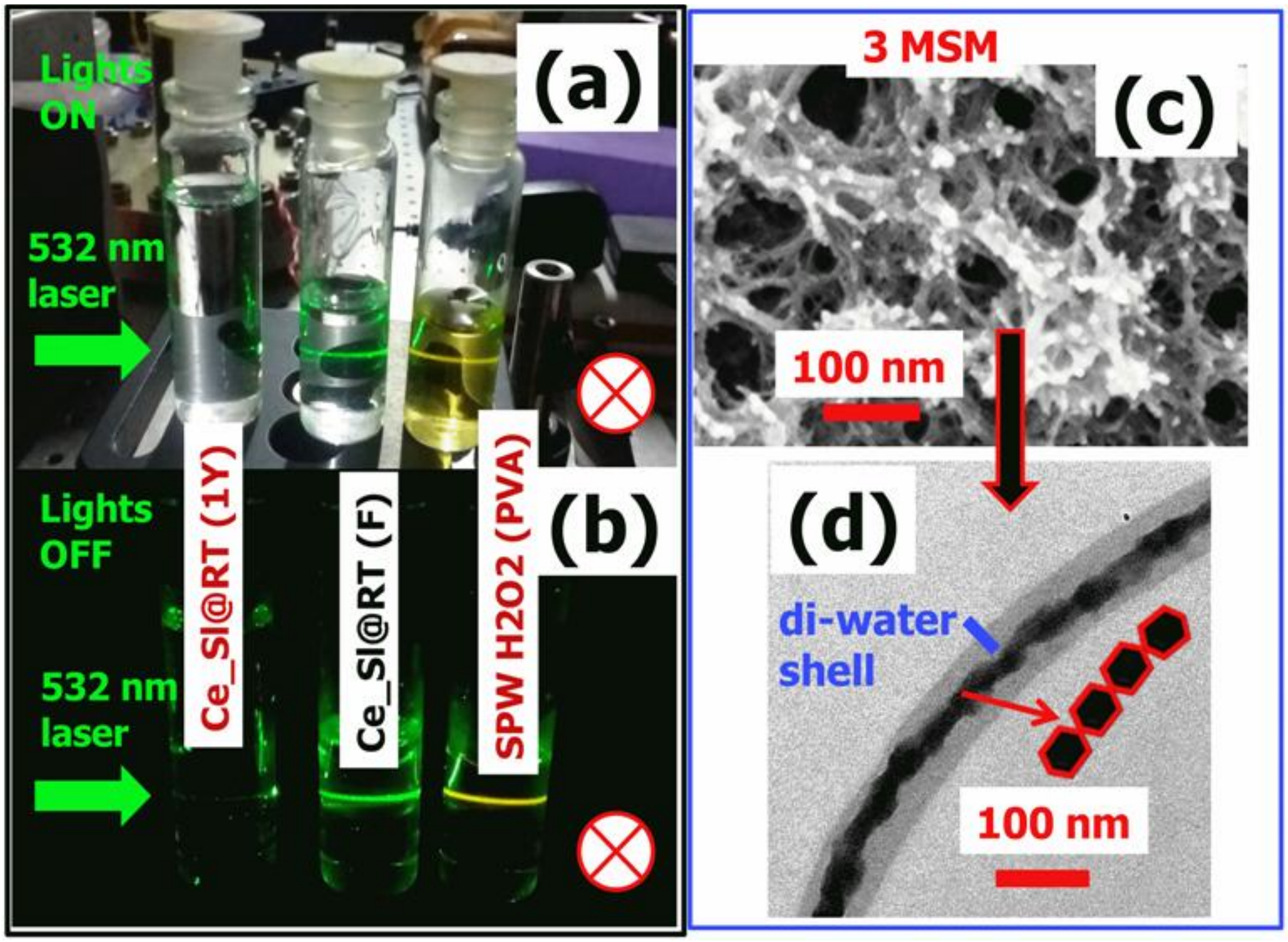

Fig.3.14 [*Light Scattering efficacy of Ce_Sl@RT product*]: Laboratory lights (a) ON, and (b) OFF state camera snaps illustrating scattering efficiency with 1year aged product. Microstructure of 3 months aged settled mass: (c) FESEM, and (d) TEM BF image respectively.

## Validation of OA mechanism in 1D-Ceria grown fibers:

The grazing incidence X-ray diffraction (GI-XRD) of the 12 MSM product drop cast on a fused silica surface is chosen to probe for 1D-Ceria fibers texture, and the structural phase analysis. The GI-XRD as a bulk characterization technique is more preferable than localized TEM investigation support for OA growth. Many TEM microstructural demonstrations already exist in literature, hence only a few localized TEM data from the 12 MSM individual fibers are acquired to supplement the GI-XRD data.

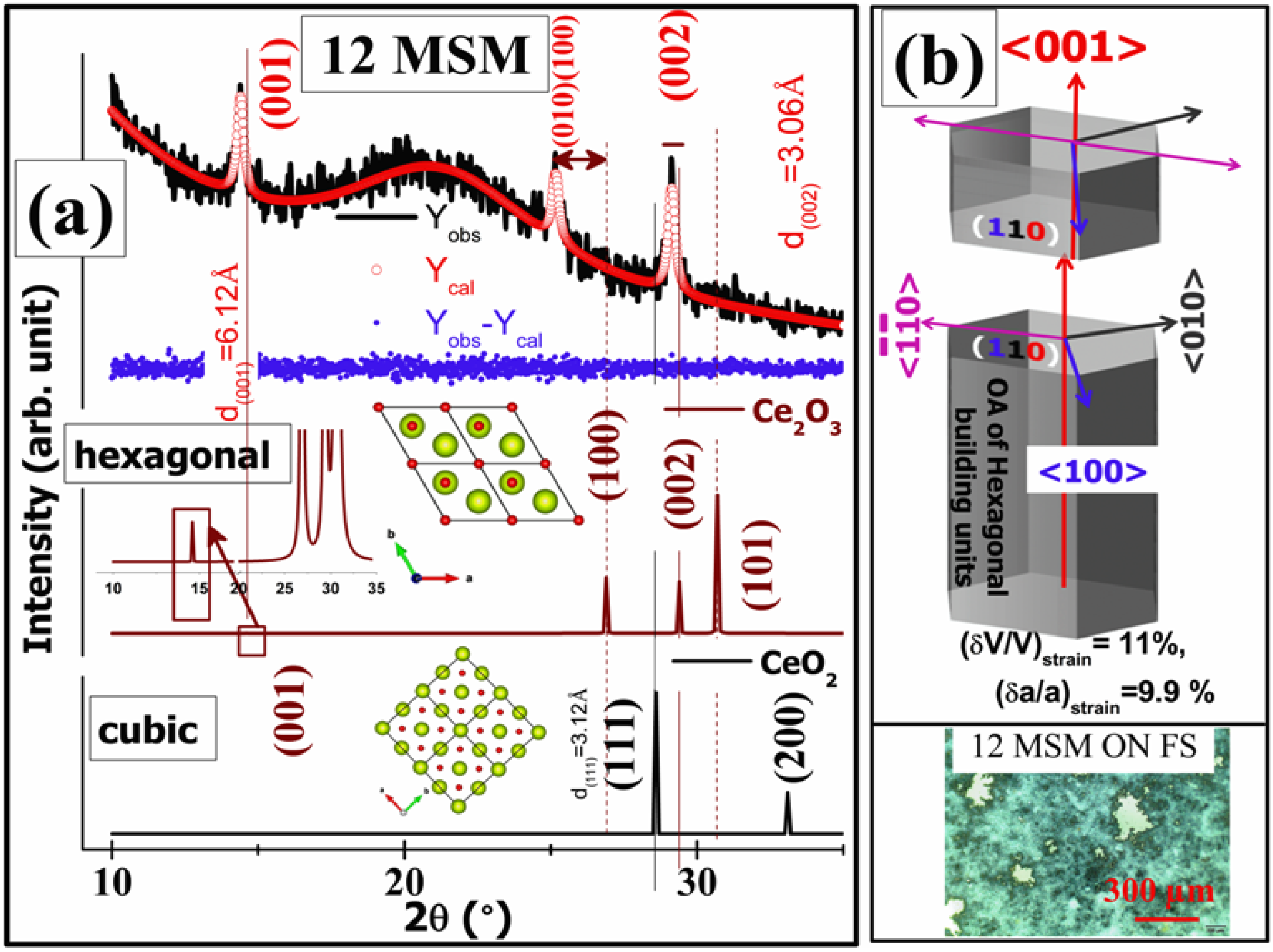

**Fig.3.15** [*GI_XRD data of 1D-Ceria fibers coated on fused silica substrates*]: (a) 1D anisotropy induced cubic ($CeO_2$) to hexagonal ($Ce_2O_3$) structural phase transition, and (b) schematic of the two hexagonal building units along (110) crystal plane OA, in accordance with the 1D-fiber obtained texture respectively.

The differentiation of near equal values of cubic $CeO_2$ most intense peak at $d_{(111)}$=3.12 Å from hexagonal $Ce_2O_3$ $d_{(002)}$=3.06 Å as the growth orientation is possible by XRD. However, in TEM for such ambiguity, the use of the SAED and HRTEM mode in a sequence is employed. The GI-XRD data of the 12 MSM product and the appropriate possible structural phases: (1) FCC $CeO_2$ (SG: 225, Fm-3m), and (2) hexagonal primitive $Ce_2O_3$ (SG: 164, p-3m1) are plotted row-wise in the fig.3.15 (a). The FPA is used to generate structural phase profile of 1D fibers $Ce_2O_3$ having <001> as the preferred orientation and also to extract crystallographic parameters [110]. The computed pattern ($Y_{cal}$) and also the difference pattern ($Y_{obs}$-$Y_{cal}$) are plotted in fig.3.15 (a) for inspection. These suggest a near convergence having the goodness of fit index ($\chi^2$=1.27) close to 1. The hexagonal structure refined unit cell parameters are: a=4.079(42) Å, c=6.121(88) Å, and V=88.2 (22) $Å^3$, highlighting that 1D fiber product is under tensile volumetric strain.

**Table-3.3:** 12 MSM product bulk X-ray structural attributes extracted

| **Sample Code** | **Crystallite size (nm)** | **Orientation factor** | **Tensile Strain** |
|---|---|---|---|
| 12 MSM | 32.6(5.8) | 0.76 | ($\delta V/V$)=11 %,<br>($\delta a/a$)=9.9 % |
| $Ce_2O_3$ | ….NA……<br>~2-5 μm | …polycrystalline…<br>$4.13 \times 10^{-4}$ | .ideal case ~ zero. |

The 12 MSM product crystal structure refined parameters of importance are listed in the table- 3.3. It can be seen that most of the volumetric strain is contained in the basal (110) plane, whereas the perpendicular c-axis direction is mostly relaxed and therefore contributes to the OA growth. A significant 5$^{th}$ order, higher orientation factor, along the c-axis justifies the dominant 1D texture developed with respect to the $\mu$–$Ce_2O_3$ ICCD PDF: 78-0484 file. Based on these GI-XRD assessments and subsequent indexing, a schematic of the OA growth consisting of two hexagonal building units is drawn for easy understanding in fig.3.15 (b). To briefly summarize, the OA reported in literature is expected to lead to the growth of a 1D-ceria structure based on two scenarios: (1) OA along {110} planes with <001> growth, and (2) OA along {211} planes having <111> as growth direction [134–141]. The present biomineralization case falls into the first growth scenario. Also, at the bottom of the fig.3.15 (b), one recoded optical micrograph of the drop cast 12 MSM product on a fused silica substrate, which is chosen for the GI-XRD investigation, is included for observation.

## **TEM Validation of OA mechanism of Hexagonal building units to develop into a 1D-ceria grown fiber:**

The pictorial schematic of the progress of 1D-ceria fibers growth and development process is presented in figs.3.16 (a)-(e). In literature, detailed TEM validations in concurrence with DFT theoretical models explain the evolutions of morphologies based on energy minimization. These specific morphologies become the basis of crystal growth and development. The specific cases of preferred exposed facets related to observed nc-ceria morphological features are in the following fashion: (1) {111} enclosed octahedrons, (2) {111} and {100} for truncated octahedrons, (3) {110} and {100} for rods, and (4) {100} for cubes [142–148]. The facets evolutions schematic in the figs.3.16 (a)-(c), are adapted from the above mentioned publications. Whereas in figs.3.16 (d)-(e) the gradual elimination of {111} planes along with <001> texture development is presented. The observation away from the <001> presents the 1D-ceria fiber having {110} and {001} planes as facets, is shown in fig.3.16 (e). However, in reality if the FCC unit cell is considered as the building unit, two important physical attributes need to be satisfied. Those are: (1) planes atomic packing surface density order should be $n_{(111)}>n_{(110)}>n_{(100)}$, and (2) their surface energy value is in the order $\gamma_{\{110\}}>\gamma_{\{100\}}>\gamma_{\{111\}}$ respectively. Based on these two physical attributes, FCC building unit derived 1D-ceria morphology should be enclosed by the {111} crystal facets. Contrary to this, the final 1D-ceria morphology reported in literature has no such faceting followed. In most of the reports the grown 1D-ceria morphology has facets with oxygen-deficiencies, leading to highly reactive facets in nature. These 1D-ceria having reactive facets have a lot of technological applications.

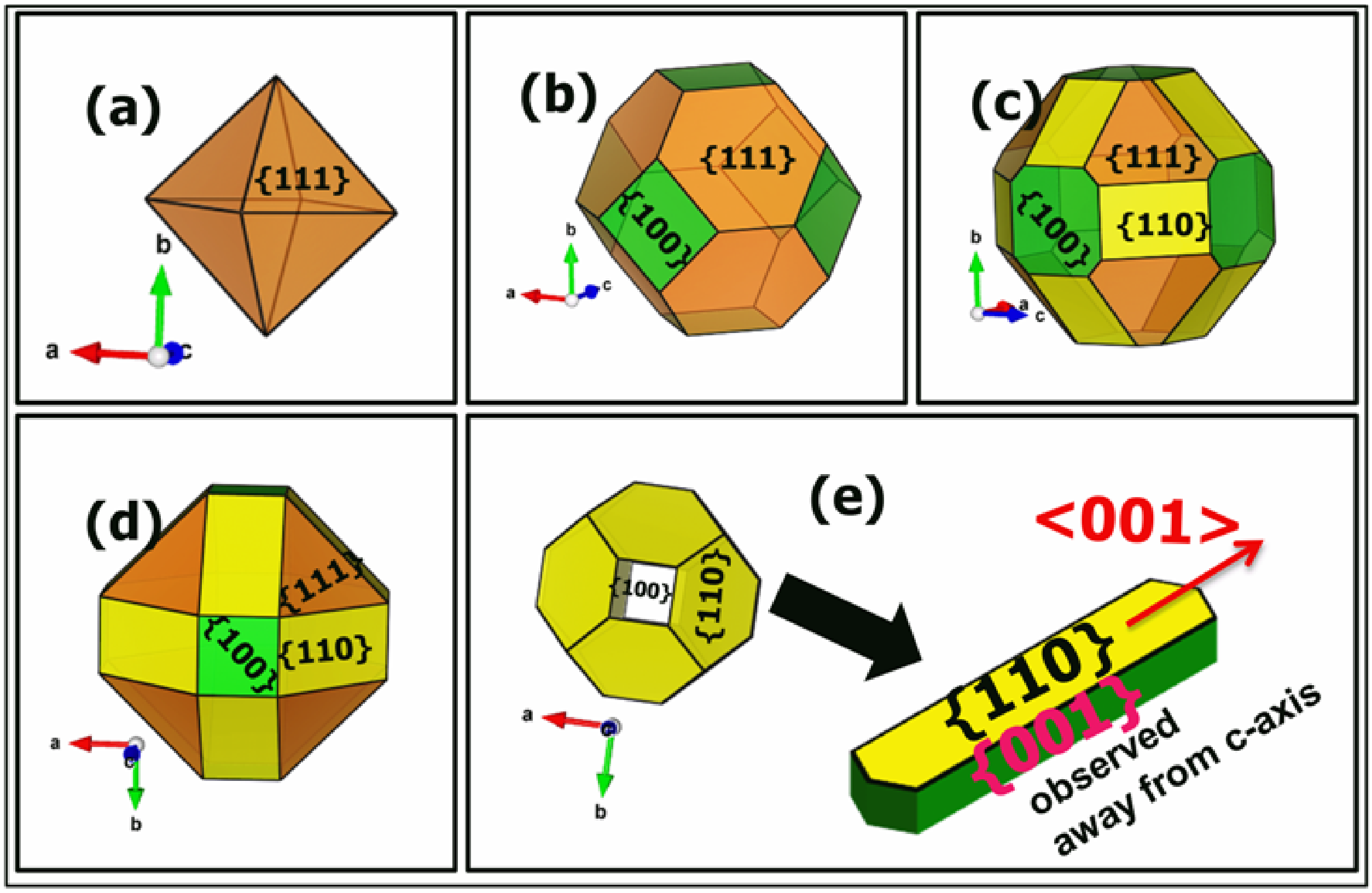

Fig.3.16 [*Growth stages of 1D-ceria fibers: Schematic view*]: Ceria unit cell; (a)-(c) depicts crystal facet development, (d) gradual elimination of {111} planes and <001> texture formation, (e) <001> textured 1D-ceria fiber observed along c-axis and away from the c-axis respectively.

This section is devoted to the localized TEM microstructural data acquired to support GI-XRD and presented inferences, leading to an insight into the OA mechanism guiding 1D nanostructures growth. It is important to identify the 1D-ceria fibers growth axis. To achieve this, a set of sequential HRTEM and SAED mode micrographs are acquired at the same localized region of the 1D-ceria fibers. Specifically, an edge portion of the fiber suitable to the TEM e-beam transparency is chosen and examined in detail. The TEM extracted microstructural and structural information with analysis from one such 1D-fiber is presented in the figs.3.17 (a)-(e).

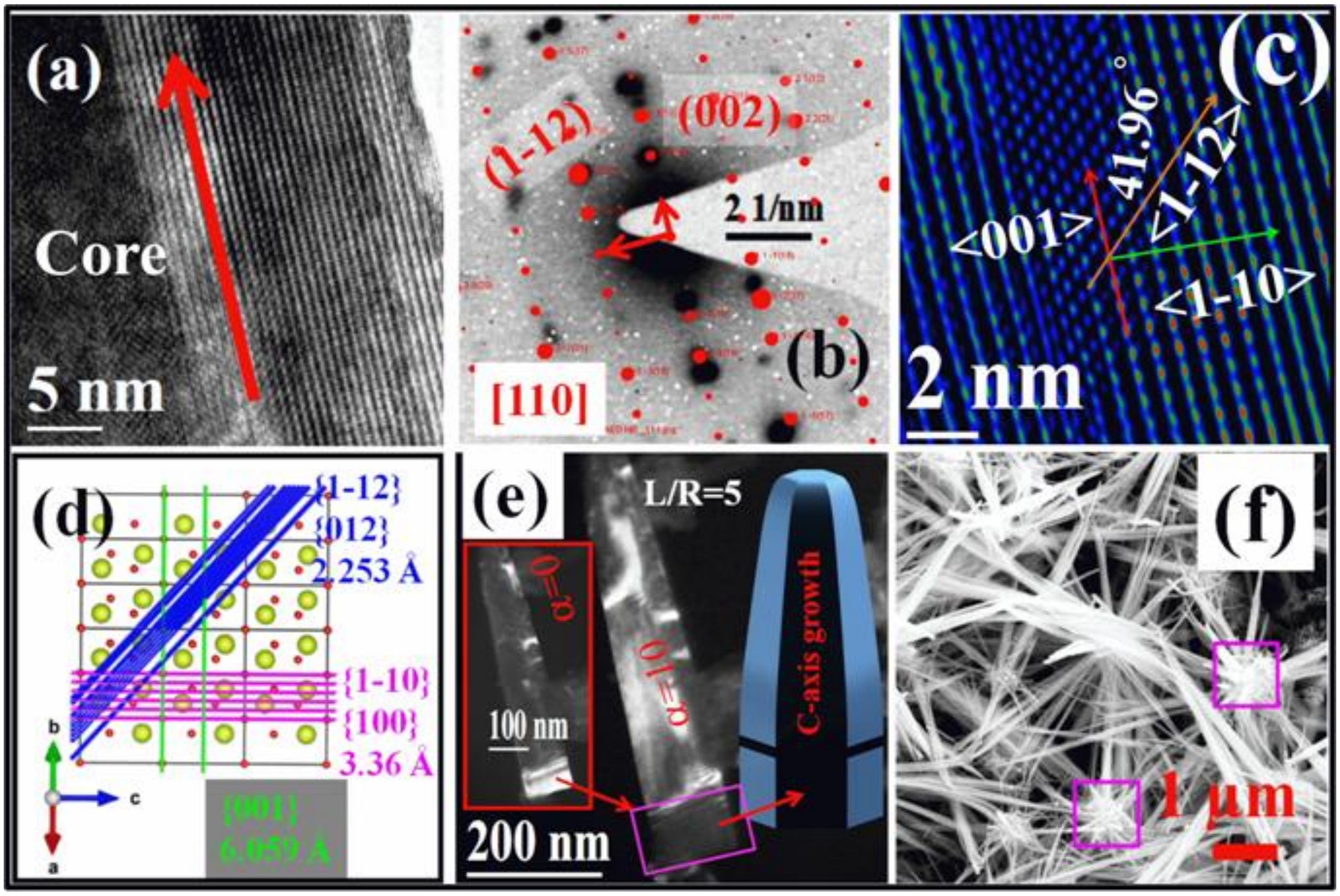

**Fig.3.17** [***Growth direction of 1D-ceria fibers by TEM***]: TEM acquired; (a) BF, (b) SAED, (c) HRTEM, and simulated HRTEM of (c) justified in (d). (e) TEM DF for α=0, 10 micrographs, and (f) FESEM micrograph of 1 D-Ceria fibers respectively.

These localized TEM microstructural observations are consistent with the bulk GI-XRD data presented in the fig.3.15, highlighting 1D-ceria fibers to be made out of a $Ce_2O_3$ hexagonal structural phase. The TEM-SAED acquired localized structural data iteratively processed for the possible zone-axis with less than 5 % tolerance, point to [110] as the preferred zone-axis. Indexing of the TEM-SAED data in this zone is then carried out. The three shortest d-spacing lattice planes of the hexagonal $Ce_2O_3$ unit cell nearest to the primary spot are evaluated and are found to be in concurrence with the XRD data. These are: (1) $d_{(002)}$= 3.03 Å, $d_{(1-12)\ or(012)}$ = 2.253 Å, $d_{(1-10)\ or\ (100)}$= 3.37 Å, and (2) m∠(1-10) & m∠(1-12)=48.04 °, m∠(1-12) & m∠(002)=41.96 °, m∠(1-10) & m∠(002)=90 ° respectively.

The one to one correspondence of the HRTEM data with $Ce_2O_3$ unit cell is realized by depicting this unit cell generated simulated HRTEM. The simulated HRTEM schematic is shown in fig.3.17 (d). The single crystallinity of these fibers in the TEM DF mode is also accessed. It is evident that the fibers are of high crystallinity but contain around 100 nm rectangular basal planes with a distinctly different orientation than the entire grown fiber portion above it. It resembles the seeded growth development scheme, in which a specific lattice-matched plane with a specific texture is facilitated to grow preferentially by the seed. It is hypothesized that an initial lumpy mass by nanoparticle aggregation gets initiated, which develops into a basal seed for the subsequent preferential growth of 1D-ceria fibers above it. The FESEM image provides direct evidence for the radial outflow of these 1D-ceria fibers from many-seed like entities. These entities are dominantly observed and are identified in fig.3.17 (f) (rectangular magenta-colored regions).

## In-situ TEM e-beam facilitated studies of OA mechanism:

## Prescribed Protocol for TEM e-beam Probe or Modification to be followed:

Both in-situ electron modification and probe activities are sequentially done in FEI Tecnai $G^2$ 20 S-TWIN TEM microscope equipped thermionic $LaB_6$ electron gun emission operating at 200 keV. A chosen specimen region is brought to focus under TEM e-beam at Mh 610 kX high-resolution mode (largest condenser aperture, 150 μm, and 200 nm condensed spot), and is irradiated with step-4 $LaB_6$ electron emission current displaced in the TEM filament supervisor console (E4I=7.93 μA, see fig.3.18 (a)), for initiating the specimen dynamic in-situ modification. A nominal step-2 $LaB_6$ reduced emission (E2P=0.88 μA, used to probe any further dynamic changes introduced) stage is employed for reloading and imaging of the radiation quenched localized modified regions of the specimen. These irradiated regions are initially fed to the stage control flap-out memory address so that retracing the exact location can be done as and when the subsequent sequential (probe or modification) operative step needs to be reactivated without any specimen location ambiguity. One such E4I specimen exposure region with beam condensed to approximately 200 nm is depicted in fig.3.18 (b), as an imprint highlighting the shape of the electron beam. It is to be noted here that the carbon support films are the standard support films in TEM because of its irradiation, thermal stability, and chemical inertness. However, various beam induced transformations such as C-atom sputtering in the beam focused region leading to thinning resulting in brighter contrast can be found in literature [149]. Beam spreading by the intensity knob and a simultaneous ~<30 s recovery to step-2 filament emission criteria recursively utilized for altering the beam attribute from material modification to probe tool. In TEMs having field emission gun ($J=10^5$ $Acm^{-2}$ current density) as electron emitter, these dynamically tracked changes will just get expedited in time scale because of three orders higher magnitude of the current density used. In contrast the current dynamical occurrences will be at slower time scale, because of having $LaB_6$ thermionic emitter investigation [48,150–153]. Significantly, even if the irradiation mode is stationary, live monitoring to retain the TEM e-beam center precisely at the specimen focused spot location is carried out concurrently as and when required. It is done to negate both the instrument specified probable specimen drift (<1 nm/min), and spot (<2 nm/min) drift respectively because of such high Mh range (approximate 25 nm × 25 nm region) data acquisitions.

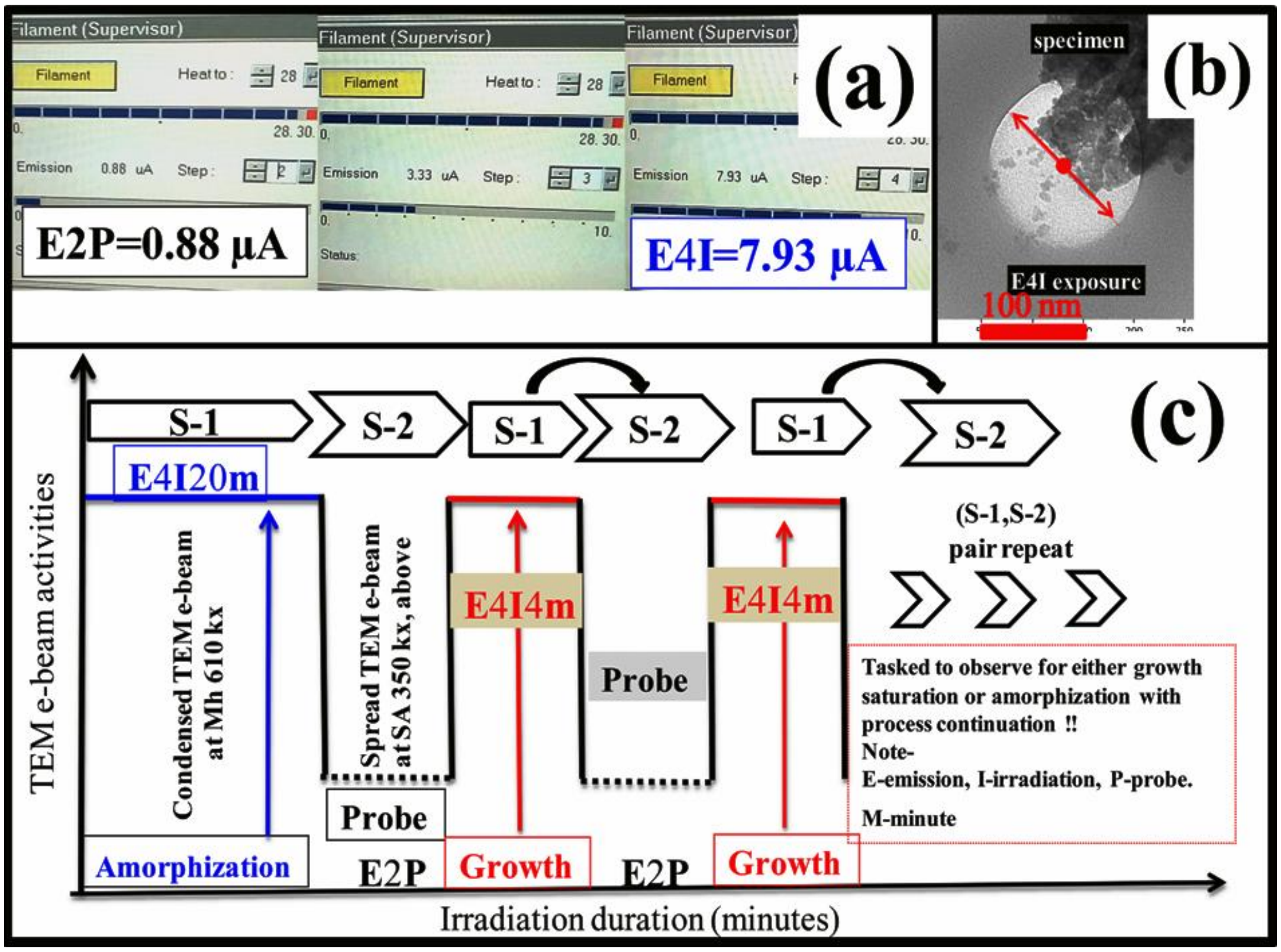

**Fig.3.18** [*Schematic of TEM e-beam activity*]: (a) $LaB_6$ filament emission to probe (E2P) and emission for irradiation (E4I), (b) E4I modified region, and (c) schematic for subsequent studies respectively.

Based on several repeated observations in the HRTEM mode an optimized combination of E2P (Emission to Probe) and E4I (Emission for Irradiation/modification) duration is quantified. The HRTEM modified region is probed employing TEM techniques like TEM BF, DF, and SAED mode etc. The optimized prescription schematic is drawn and is shown in the fig.3.18 (c). The E2P time can extend to 10-15 minutes based on retracing and acquiring micrographs, whereas E4I sharply limited to 4 minutes. The E4I4m (Emission for Irradiation maintained for 4 minutes) is mostly devised to facilitate nc-ceria growth, out of the central region of the amorphized ceria layered flake utilizing an initial 20 minutes TEM e-beam hammering [154]. The purpose of activating TEM e-beam hammering is to develop a radiation-hardened region, i.e., to have almost negligible atomic lateral movements as a result of dense packing and void removal. It will also provide an electron transparent region for the TEM HRTEM mode imaging. Based on a few experiments, it is expected that a 20-minute electron hardening of the localized region will be able to sustain the same emission dose for at least a further subsequent period of 20 minutes without any degradation. As a result of this, the total 20 minutes irradiation/modification duration is further sub-divided into 5 divisions for the quantitative growth kinetics extraction. Incidentally, the optimized E4I4m exposure produces HRTEM readable crystal lattice growth for all the subsequent TEM mode investigations.

## TEM e-beam devised Hammering and Amorphization (E4I20m):

The TEM BF, HRTEM, and SAED mode are employed cooperatively to investigate many TEM e-beam hammered nc-ceria lump irradiated locations for "radiation hardening." The "radiation hardening" alludes to a localized TEM e-beam transparent; atomically dense (devoid of any pores, voids, or gaps) packed amorphized ceria thinned flat regions. As a result this region will have negligible lateral movement (radial expansion perpendicular to the e-beam direction) during the irradiation facilitated subsequent crystal growth investigations. The microstructurally distinct two regions of nc-ceria formulations that are observed in TEM BF micrographs are shown in fig.3.19 (a)-(b). VCX 750 W 20 kHz ultrasonic processor with a 13 mm solid ultrasonic horn and operated at 50 % of its maximum amplitude for 5 minutes, is employed to obtain nc-ceria dispersions in 25 mL of DI-water. A pinch of this nc-ceria powder dispersion is capillary drop cast over a TEM grid and the obtained nanoparticulate well-dispersed form in TEM BF mode is shown in fig.3.19 (a). A statistical analysis of 5-6 such TEM BF micrographs over 100 particles suggests to the lognormal distribution, with average nanocrystal size, is of 1.9(0.4) nm.

Similarly, the TEM BF images of nc-ceria lumps dispersed in a water-based ultrasonic cleaner (Ralsonics model R-80 W, 30 kHz) for 5 minutes are recorded. These micrographs indicate that the sample has crystallites with random orientation and nearly densely arranged as seen in the fig.3.19 (b), and its representative drawn schematic respectively. The presence of voids, pores, gaps, and amorphous ceria matrix regions connecting nanocrystallites is also observed. These nc-ceria lump formulation having nc-ceria crystallites in contact with the adjacent ones through crystal boundaries and ceria amorphous matrix is the specimen of choice for subsequent investigations. As presented in the previous section, E4I TEM e-beam focused onto the lump-edge gradually moved towards thick lump center at Mh 610 kX. It enables the lumps to flatten and develop into atomically thin fused electron transparent region of almost 25×25 nm dimension which is the minimum requirement for probing (see HRTEM micrograph fig.3.20 (a)). The E4I hammering for thinning of the nc-ceria lump region is continued for 20 minutes (E4I20m), and the micrograph shown in the fig.3.20 (a) suggests that the optimized time span is sufficient for fabricating ~ 25(+5) sq. nm radiation-hardened amorphized lamellar ceria matrix locally. This operation (E4I hammering for 20 minutes) is tracked over many such aggregated lumps, and one such TEM BF colored micrograph is presented in the fig.3.20 (b). To have a clear, distinct view (fig.3.20 (b)), the TEM BF micrograph is color indexed. It is carried out by the digital micrograph "histogram palette" script to adjust colors. The rule being that the micrograph pixels with a specific range of intensities in the image are colored as one entity. Thereby, the amorphous carbon support film (C-film), TEM e-beam hammered, and the pristine nc-ceria locations of the fig.3.20 (b) are colored to green, dark-gray, and white individually for illustration purpose, respectively.

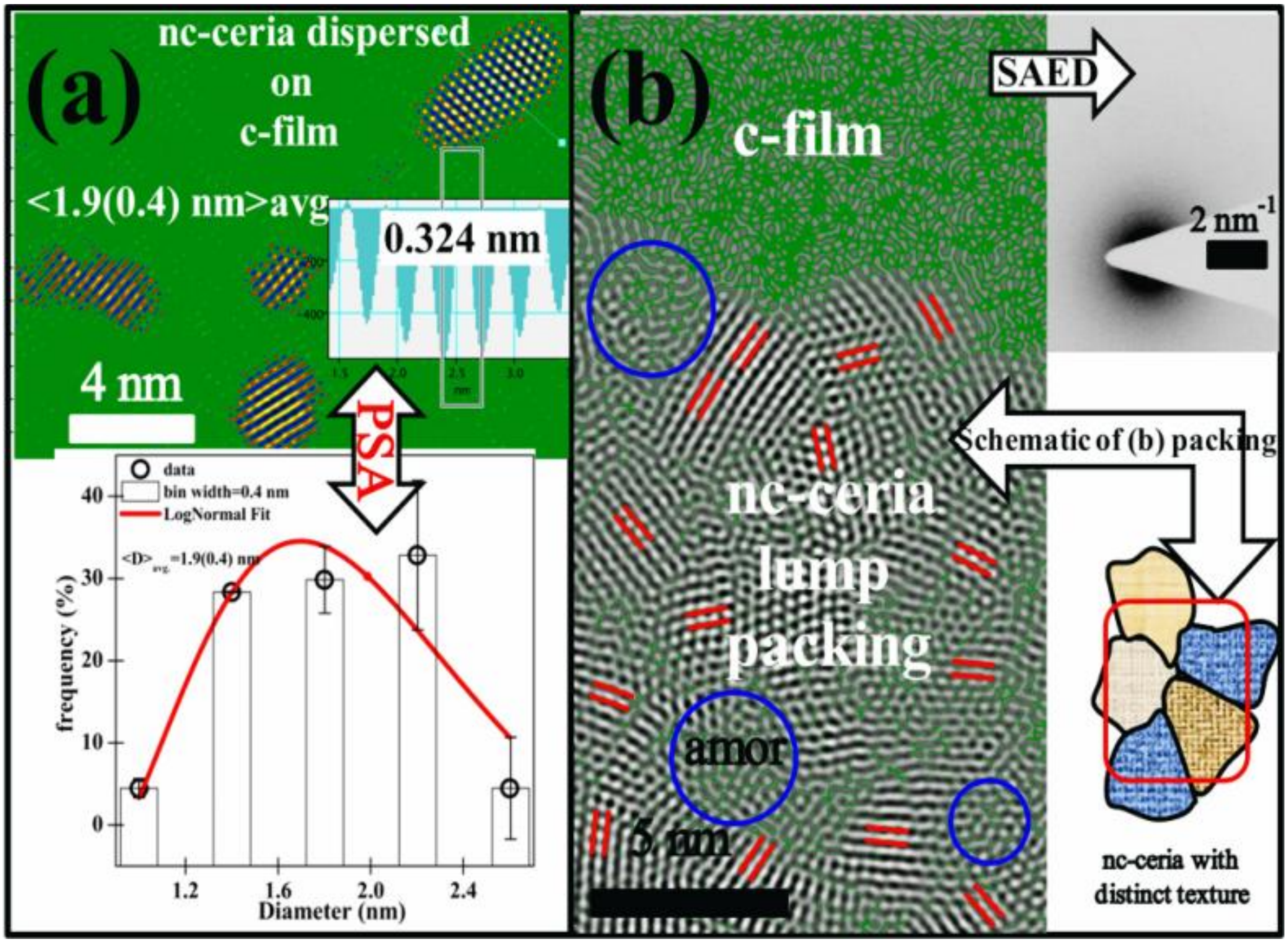

**Fig.3.19** [*Micrographs of two distinct nc-ceria formulations*]: Ultrasonic (a) horn, (b) cleaning bath dispersed nc-ceria HRTEM micrographs respectively.

To ascertain the, TEM e-beam hammered morphological change leading to the amorphization process, TEM SAED aperture is introduced. The region-specific diffraction patterns are recorded. In fig.3.20 (c) a comparison of two such regions recorded diffraction patterns of the nc-ceria before and after TEM e-beam hammering is shown. Both the diffraction patterns are overlaid with respect to each other, for which Al $d_{(111)}$= 0.4277 $Å^{-1}$ is considered as the calibrating Debye ring. The nc-ceria region TEM SAED digital pattern is matched directly with the PROJECT/PCED2s crystal structure generated simulated pattern. By using the simulated pattern the recorded pattern spacing are extracted and are labeled as shown in the fig.3.20 (c) [155]. The ring radii and the corresponding intensity level ratios are transformed into a simulated linear electron diffraction pattern (LDP). It has a clear match to that of stoichiometric cubic ceria structural phase. In the TEM e-beam amorphized region, the intensity of these Debye bottom half rings is almost nullified (zero contrast variation) after the first disc, validating TEM e-beam hammering induced amorphization. The distinction being that TEM e-beam hammering generated microstructure is dominated by a densely fused amorphous ceria matrix, having sparsely populated ceria nanocrystallites. All these ceria nanocrystals are seen to be of the same orientation (incidentally mesocrystal like appearance), suggesting that with an appropriate TEM e-beam excitation the surrounding disordered ceria matrix can be consumed, facilitating the subsequent growth of ceria nanocrystals.

The list of possible phenomena under the TEM e-beam are (1) mass loss, (2) valency state reduction, (3) phase decomposition, (4) precipitation, (5) gas bubble formation, (6) phase transformation, (7) amorphization, and (7) crystallization. These attributes, in the case of the oxides in TEM, are already reviewed and presented elsewhere [156]. Here this literature is scrutinized to present a viable interpretation of the amorphization process. Assuming the TEM e-beam operation is confined to 200-1000 keV probe energy range, based on elastic collision, the maximum transferable energy must be 4-31 eV for Ce, and 33-269 eV for O ions respectively [157].

Similarly, the computed assumed displacement threshold energy of Ce-atom in ceria sub-lattice is 40 eV; likewise, for O-atom, it is 20 eV constituting the oxygen sub-lattice of the fluorite structured ceria [158].

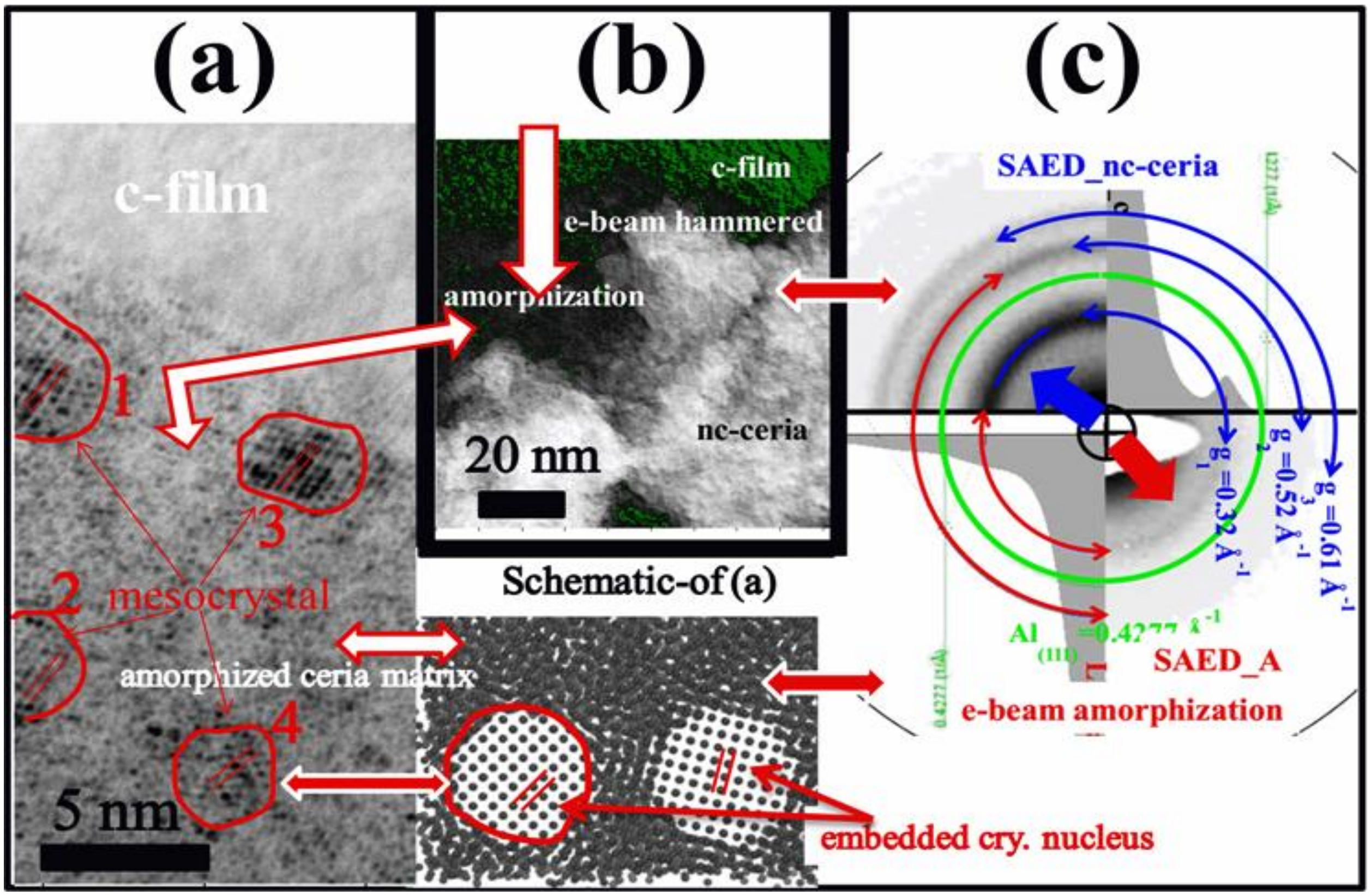

**Fig.3.20** [***Micrographs of TEM e-beam hammered amorphized region***]: (a) nc-ceria lump with distinct e-beam hammered region, (b) HRTEM, and (c) comparative SAED patterns of nc-ceria and amorphized region respectively.

A comparison of these two energy values (i.e., maximum probable transferable energy, and displacement threshold energy) with TEM 200 keV e-beam operation suggest that generation of O-atoms knock–on displacements (non-thermal amorphization process) is the most probable process in nc-ceria fluorite structured oxygen sub-lattice. Another exciting facet of this fluorite structured ceria lattice is that these O-atoms associated defects are observed to be actively annihilated even through the TEM microscope in ultra-high vacuum environment [159]. It will lead to O-defect creation and annihilation concurrently. Thus, the rule will be that if defects creation overtakes the recovery rate then, defects accumulation (point defects or chemical disorder) will substantially lead to the initiation of the localized topological disorder. This is consistent with an earlier study of Bhatta et al., who reported amorphization of 15 nm diameter nc-ceria under 200 keV TEM e-beam [160].

Also, nc-ceria as a facilitator of amorphization is reported in literature [161]. It is suggested that presence of nc-ceria in conjugation at the oxide heterointerfaces locally enhances the radiation-induced amorphization process of the adjoining oxide. Moreover, two primary models proposed for justifying the amorphization process under the TEM e-beam irradiation are (a) the mechanical thrust of the bombarding electrons delivered to the target atomic nuclei, i.e., knock-on mechanism; (b) electronic excitation by the electric field, i.e., radiolytic process [162]. The physical understanding of the amorphization and its corresponding reasoning, with a list of few specific materials investigated earlier includes: (a) electronic excitation in-titanate pyrochlores (anion disorder in $A_2Ti_2O_7$, A=Y, Gd, and Sm), colloidal $CsPbBr_3$ nanocrystals, complex rare earth (Y(III), Eu(III), and La(III)) nanostructures etc., whereas (b) the frequently encountered knock-on displacements mechanism are observed in-SiC respectively [163–166]. More recently, fast and reversible phase transitions (amorphous to crystalline) have been observed in the case of the chalcogenide phase-change material (i.e., on $GeSb_2Te_4$ thin films) examined under TEM e-beam irradiation [167].

Taking a hint from reported data, the probable dominant mechanism responsible for the amorphization observed in the nc-ceria is attributed to the knock-on displacement of the O-atoms. Under continuous TEM e-beam exposure, the accumulation of these knock-on displaced O-atoms is practically feasible. Thus, the 20 minute continuous exposure accumulates enough O-atom point defects, which locally suppress the nc-ceria crystalline structure resulting in localized amorphization. The amorphization region seems to cover the entire TEM e-beam exposed regions. Based on these inferences, the amorphized ceria matrix region can be identified as oxygen-deficient non-stoichiometric ceria fraction (see fig.3.20 (a)).

Significantly a sharp time limit separating the; (1) first complete amorphization process, and (2) second initiation of the crystallization step is not easy to distinguish. It can be seen in the micrographs, both these processes overlap. The nucleation of a few embedded stoichiometric nc-ceria nuclei developing into a mesocrystalline structure starts evolving. These nucleated nc-ceria crystallites gradually consume the surrounding atomically disordered ceria matrix fraction to grow further. The periodic TEM e-beam excitement directly facilitates the growth. In order to present a comparative view of the amorphized process carried out under the TEM e-beam, two regions are shown in fig.3.20 (a)-(b). The region identified in fig.3.20 (a) is an amorphized region, whereas fig.3.20 (b) is that of the corresponding pristine nc-ceria region respectively. Randomization in the context of TEM-SAED refers to the availability of the minimum crystalline fraction available to contribute in generating localized diffraction from the probed nanometric region.

## Probing the TEM e-beam Grown Crystalline Region (E4I16m):

Localized growth under TEM e-beam from a nc-ceria lump is another possibility contrary to amorphization, which needs selected area structural investigation. The "region 1" marked in the fig.3.21 (a) represents TEM e-beam localized irradiation grown 18-21 nm crystallite (based on the prescription presented in the previous section, see fig.3.18 (c)) surrounded by irradiation unaffected nc-ceria lump "region 2". Typically, a TEM-SAED aperture inserted to the objective lens image plane with the specimen region of the importance of at least 15 nm in size can be inspected employing the TEM-SAED mode to extract localized structural information [168,169]. As expected, this localized TEM e-beam grown region is single-crystalline, with a typical spot diffraction pattern, as shown in fig.3.21 (b).

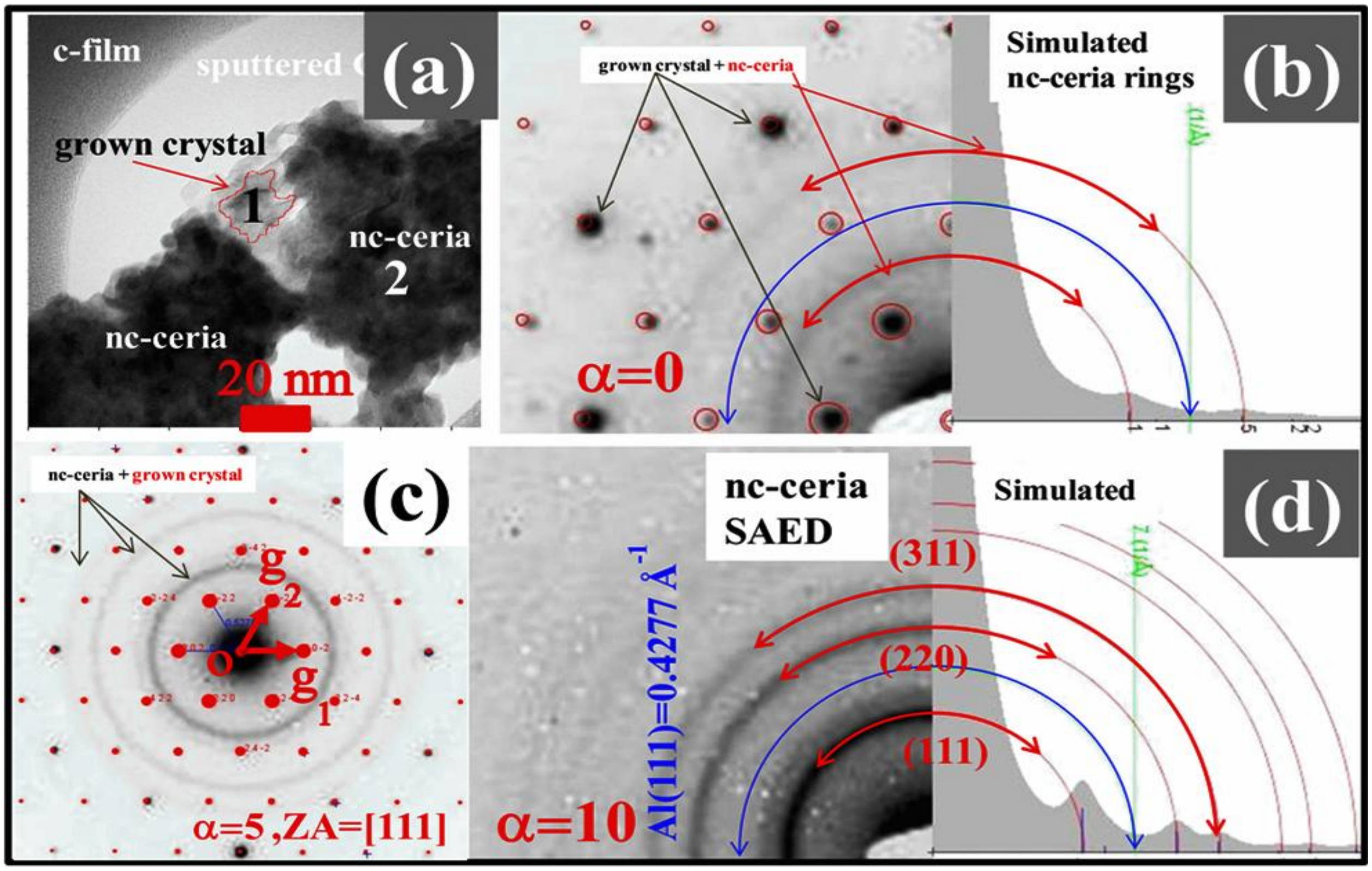

**Fig.3.21** [***Probing e-beam grown nano-crystallite***]: (a) TEM BF micrograph of the e-beam grown nanocrystalline (marked as 1), and the original nc-ceria region (marked as 2) respectively. (b) α=0, (c) α=5, and (d) α=10 e-beam tilt series acquired SAED patterns acquired from (a) with the aperture centered on grown crystallite.

The surrounding region as diffuse rings present along with the highly diffracting spots pattern of grown nanocrystalline "region 1". The nanocrystalline region obtained intense spot pattern is shown in the fig.3.21 (b). Subsequent beam tilting TEM-SAED recorded from the same region can be seen in figs.3.21 (c) and (d). It improves the polycrystalline ring patterns of the surrounding nc-ceria region, which becomes dominantly observed for α=10 tilt. Possible zone-axis search by one to one reciprocal lattice matching is carried out with the simulated spot-patterns generated employing PROJECT/SAED2s [170]. The ceria structural data (ICCD PDF: 81-0792) is utilized to generate the simulated patterns, and the most appropriate pattern with a matching tolerance of less than 5 % is chosen. The simulated pattern of [111] zone-axis matches well the experimental TEM-SAED, with the reciprocal lattice mismatch of about $g_{(2-20)}$=1%, $g_{(20-2)}$=1%, and in-between angle are of $m\angle g_{(2-20)}$ O $g_{(20-2)}$=0.5% respectively.

Similarly, the nc-ceria fingerprint electron diffraction digital ring pattern is matched directly with the PROJECT/PCED2s crystal structure generated simulated pattern, which is shown in the fig.3.21 (d) [155]. The ring radii and the corresponding intensity level in the simulated linear electron diffraction pattern matched well, thus validating the structural phase to be that of the chemically ordered cubic ceria phase. Hence, the ability of the TEM e-beam in delivering localized nanocrystalline orderly region out of a fully radiation-hardened amorphous localized portion is demonstrated.

## TEM e-beam Facilitated Aggregative Growth and Growth Kinetics:

After fabricating the localized amorphization (continuous E4I20m), and the subsequent in-situ mode tracking of the amorphized region for NCs growth (regular E4I4m pulses having an intermittent gap) is pivotal to understand growth kinetics. Many amorphized regions are considered, the best one having utmost digital clarity is presented for observation in the fig.3.22. To compare the current experimentation using the pulsed TEM e-beam stimulus, the case study of metal NPs growth employing pulsed potential conditions is considered [171]. The point being that, in the presence of this pulsed electron stimulus (E4I4m), the fabricated amorphized ceria matrix region atoms get activated for atomic transport. This induces localized atomic order (atom cluster formation) with the possibility of simultaneous oxygen vacancy healing leading to many randomly nucleated crystalline nuclei. In the classical term, this amorphous ceria matrix embedded crystalline nuclei would consume the available surrounding randomized atoms for growth. In contrast, the non-classical case will be the co-operative interaction between the neighboring crystalline nuclei leading to the growth. A densely packed TEM e-beam randomized nc-ceria nuclei region is shown in the fig.3.22 (a). The aim of amorphization is to eliminate previously present intra-particulate voids. However, in the amorphized micrograph of the fig.3.22 (a), these observed dark contrast regions are mostly the oxygen defects or oxygen cluster defects. These will be healed as and when TEM e-beam changes from the material modification to probe attribute, and thereby facilitating recovery. The possibility of cationic surface reconstruction triggering during the TEM e-beam exposure also cannot be ruled out [172]. The explanation of why an amorphous, disordered structure (having relatively higher internal energy) will undergo energy lowering to develop into a crystalline (highly ordered) even in the presence of a constant TEM e-beam energy input impulse can be found elsewhere [154,173,174]. In addition, there are reports of in-situ liquid cell TEM tracking of two adjacent distinct NPs coalescence through OA driven non-classical crystallization growth is plenty [175–178].

In the present context, one of the larger grown morphology demarcated crystallite is shown in the figs.3.22 (a)-(f). Its time evolved quantitative growth after consuming adjacent primary crystallites like nutrients following the prescribed TEM e-beam protocol (see section 1) is recorded. The (nano) crystal size distribution (CSD) of t=0 minutes micrograph is plotted in the fig.3.22 (c) (as inset is bimodal). The smaller distribution corresponds to primary nanocrystals, and the broader just initiated distribution represents aggregates of the primary. Similar sets of experimentation are extended to almost 5 other regions for the time-evolved crystal growth data collections to present a quantitatively accurate assessment of the facilitated growth kinetics. This quantitative data is expected to lead to provide insight into conditions where OA is the sole dominant mechanism directing the observed crystal growth. By starting with an initially preformed densely packed crystal distribution (see the inset of fig.3.23 (c)), the usual classical growth process is eliminated. The remnant aspect only is to distinguish between the aggregative growth (AG) and the Ostwald ripening (OR) mechanism. This approach, as presented previously, is also used in the present investigation [171].

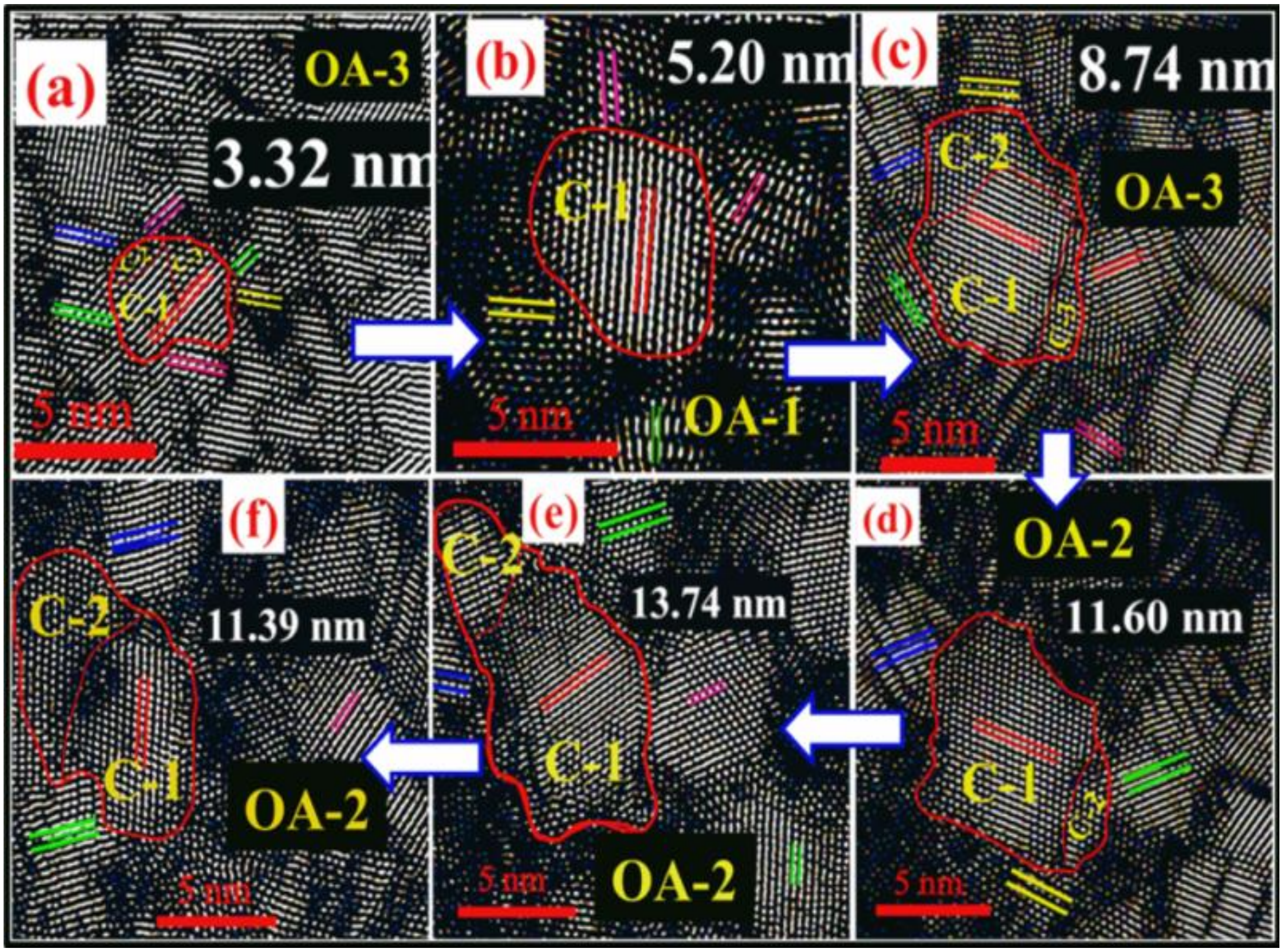

**Fig.3.22** [***Probe region TEM e-beam irradiation hardened***]: Sequential time evolved BF images depicting a single nano-crystal growth form; (a) amorphized (t=0), (b) t=4, (c) t=8, (d) t=12, and (e) t=16 minutes of step-4 e-beam irradiation. Images for subsequent longer exposure snapped micrographs are in (f) t=20, and (g) t=24 minutes (not shown here) respectively.

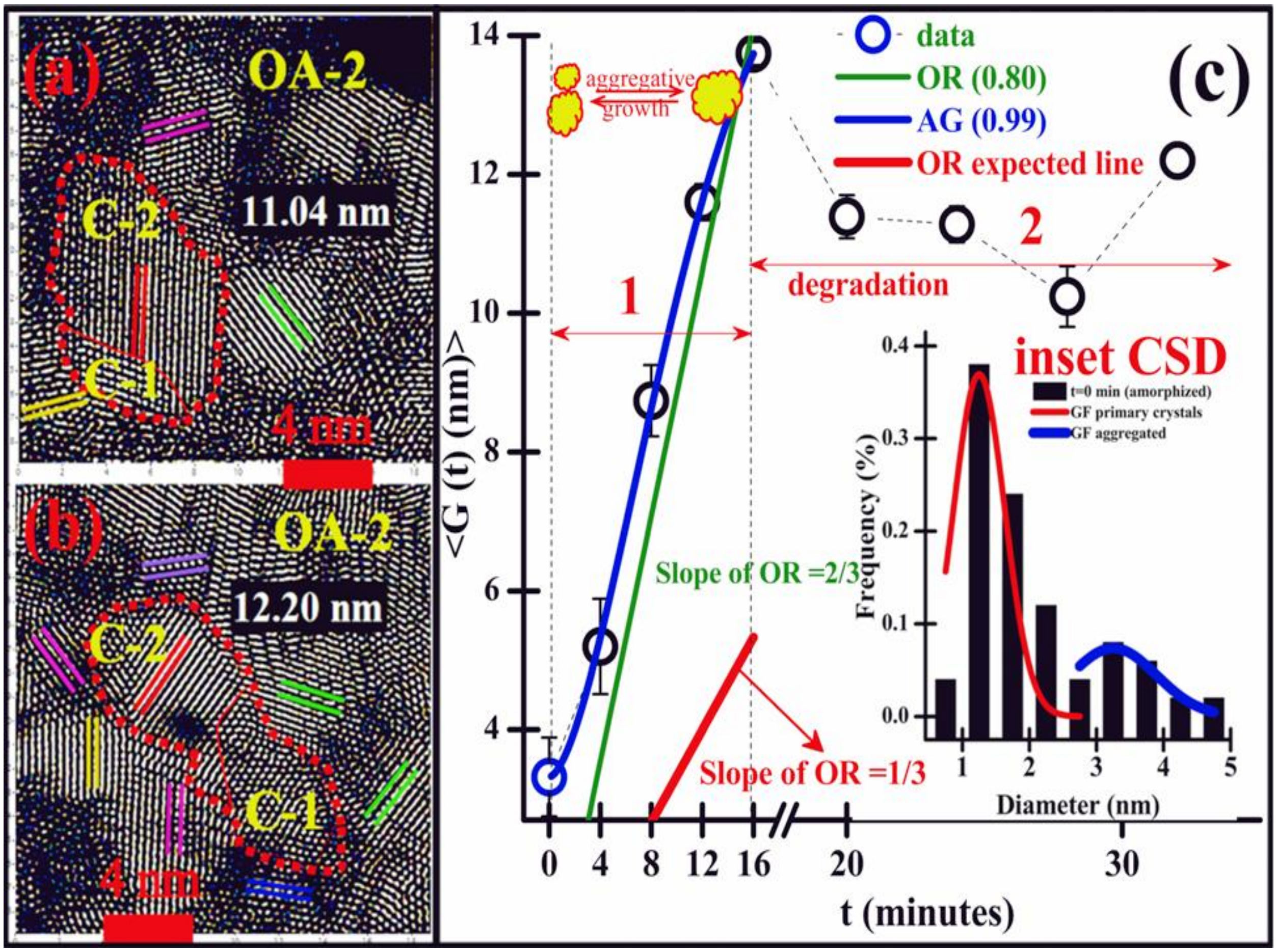

**Fig.3.23** [***Probe Region TEM e-beam irradiation Hardened***]: Sequential time evolved micrographs for duration; (a) t=28, and (b) t=32 minutes respectively; (c) is the representative growth kinetics derived with the inset corresponding to CSD of t=0 min TEM e-beam amorphized region.

The averaged time evolved growth data (<G (t)>) is plotted, with TEM e-beam irradiation time (t, in minutes) as shown in the fig.3.23 (c). The maximum time evolved single-crystal dimension in the presence of the TEM e-beam is noted and is ascribed as the growth term (G (t)). These time-evolved micrograph snapshots (figs.3.22-3.23) are insufficient to justify with clarity about the mechanistic occurrences under TEM e-beam. Given this for growth identification, the collected quantitative data is best suited to distinguish between the AG from the OR mechanism.

An iterative fit to the avg. growth data by AG and OR models signify that AG (best fit with sigmoidal kinetics) is the mechanistic happening driving growth under TEM e-beam excitation. The attempted linear fit (of slope 2/3) representing the OR mechanism is unsatisfactory, since the expected slope is about 1/3 as reported for the case of nano-Ag growth [176,179,180]. This TEM e-beam facilitated AG through oriented attachment (OA) mechanism of primary crystallites by rotation is the predominant mechanism of crystal growth observed in nature.

## OA Crystal Growth Facilitated by TEM e-beam Irradiation:

The dynamical evolution of in-situ TEM e-beam facilitated OA crystal growth is investigated utilizing the HRTEM BF imaging mode. In order to have conclusive evidence previously optimized and fabricated ceria flake of (approximate 24.8 nm × 24.7 nm dimension recorded at and above 610 kX magnification) is modified and probed alternatively. In this context, TEM e-beam dual attribute for material modification (when operated with step-4 $LaB_6$ filament emission, ~10.08 μA) and also as a probing (sequential operation at step-2, ~0.88 μA) tool is employed as per the requirement [51]. Immediately after modification duration is over, instant quenching is achieved by the compustage Goniometer α=15 tilt with a simultaneous recovery of ~<30 s. This probe region (x, y) position, including tilt, is stored in the stage setting for accurate return to the same location with absolute certainty. A sequence of modification and quenching pairs are repeated over as many as 4-5 cycles to acquire data sets representing the actual dynamical occurrences by recording the HRTEM images. The acquired HRTEM images are further processed with the Gatan's Digital Micrograph to extract inferences. Also these digital micrographs demonstrate the unit or sub-unit cell length tracked ongoing crystalline changes that are associated with the OA driven crystal growth formalism.

For this illustration, three incoherently aligned nanocrystallites (of about ~4-6 nm) but intimately fused crystalline region is chosen to track the OA growth and is shown in the fig.3.24 (a). The step-4 radiation-hardened ceria flake central region is selected for this task. Here, the identified nanocrystallites are in direct contact to restructure the conventional defective boundaries between them without being ruptured apart (i.e., free to rotate but confined for the lateral movement). The physical impetus for this type of crystal growth can be ascribed to the difference in excess energy of atoms at the crystal boundary (~1.08 $J/m^2$) compared with that of interior atoms (0.77 $J/m^2$). This process propels the boundary atoms to readjust with the immediate neighbouring crystal interior enabling growth [181]. The data presented provides clear dynamical insight into the mechanistic at nanoscale leading growth by particle-particle attachment scheme. This is unique, since previous reports focus mainly on liquid cell TEM observations in which the solution environment facilitates the OA but TEM e-beam is actively engaged for probing purpose only [45–47].

The time evolved snapshots of the TEM e-beam facilitated non-classical pathway crystallite growth process imaged for demonstration, interpretation, and inferences respectively are presented in figs.3.24 (b)-(e). As stated earlier, a fused grain formed out of the three crystallites (4-6 nm) without any core defects is shown in the fig.3.24 (a). However, the crystalline boundaries enclosing these crystallites are embedded with defects as a result of inbuilt misorientations. This large angle misfit (initial misfit of 20 ° in both <022>, <200> directions) between crystal-1 and -3 induces twinning at the adjoining boundary (dislocation lines with two regions; regions of proper fit and regions of poor fit can be seen). In contrast, the vacancies and the edge dislocations constitute stress release centers at the other two crystal boundaries. All together these crystallites constitute a central tri-junction in this micrograph. The thickness of these crystal boundaries is of order of ~1 nm and is an extension of crystal periodicity rather than an amorphous fraction. All three nanocrystals are viewed along the <011> direction.

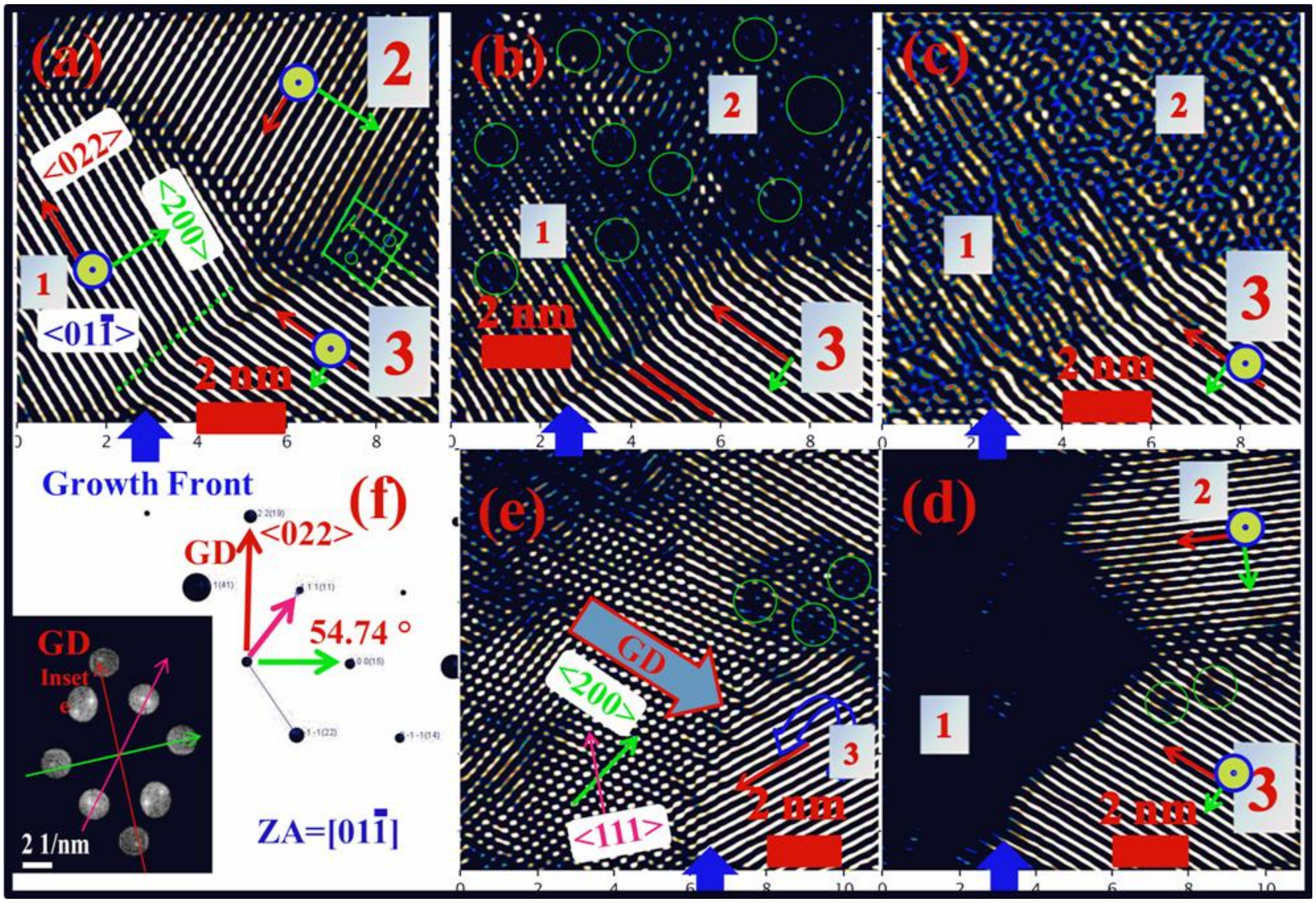

Fig.3.24 [*Probing the TEM e-beam Radiation Hardened Region*]: Sequential in-situ HR-TEM time evolved BF images after; (a) t=0, (b) t=4, (c) t=8, (d) t=12, and (e) t=16 minutes of step-4 e-beam fluence irradiation. Evolved nano-crystallite tracked in (e), growth direction is evident in (f).

The following 4 minutes of step-4 TEM e-beam irradiation-induced snapshot (see fig.3.24 (b)) elucidates the development of the crystal-1 and -2 into a single larger crystal filled with core vacancies (circular markings on the micrograph). Significantly, during this duration crystal-3 remains unperturbed. The reason for the crystal-3 being unperturbed up to 3 cycles irradiation exposure progress is the <011> directional attachment, which was previously investigated for imperfectly attached PbTe (FCC) nanocrystals [182]. A clear indication of the two diverging events that are (1) radiation-induced in-between boundary disappearance (crystal-1 and -2) and (2) bond breakage leading to vacancies generation. These are snapped in the micrograph of the fig.3.24 (b). The present observation is in conformity with the case study of multitwinned $Zn_2TiO_4$ (FCC) nanowire examined under TEM e-beam over prolonged (28 minutes) period. In that study healing of the boundary dislocations to deliver larger single-crystalline subunits by fusing OA attached nanocrystallites was observed [183]. Similar to the current dislocations, healing investigation under irradiation (3 MeV $Au^+$ ions) also validates that even an asymmetric nc-ceria crystal boundary can atomically rearrange itself to become symmetric, thereby lowering system energy by diffusing oxygen vacancies to the free surface [184]. It is also an investigated fact that nc-ceria possesses unique self-healing response to radiation damage at the crystal boundaries (volume fraction of interface region to the crystalline area for a 6 nm crystallite ~32 %) to control the crystallite size at the nanoscale [185]. Lastly, under irradiation, crystal growth is not asserted to be a thermal event, but irradiation-induced defect-stimulated mechanism at room temperature, with the primary defect being the oxygen vacancies [186,187].

Furthermore, theoretical computation of parameters like activation energy and diffusion coefficients for cerium and oxygen atoms in ceria lattice demonstrates extreme stability of cerium in both the interstitial and vacancy formulation. In contrast, oxygen in the interstitial position is metastable and also is highly mobile as a vacancy [188]. In which as that of the present 200 keV incident TEM e-beam energy interstitial-type non-stoichiometric dislocation loops of the oxygen platelets are also observed. In contrast, at least 1250 keV is the critical TEM e-beam energy that is necessary to induce cerium atom displacement (interstitial-type perfect dislocation loop) in ceria lattice [189]. Based on these literature reports, the missing atoms in the core region of the larger grown crystallite of the fig.3.24 (b) can undoubtedly be ascribed to the oxygen vacancies. Ceria in its nanostructured form (nanocubes ~30 nm) under a 300 keV TEM e-beam exposure suggests reduction (oxygen vacancies creation) and also the simultaneous annihilation in vacuum [190]. It is stated that oxygen atoms of the order of $10^6$ are available within the ceria lattice. In contrast, within the TEM sample chamber there is oxygen molecules of the order of $10^{13}$ available to impinge into the ceria lattice. The magnitude of this differential strength in the number of oxygen molecules is sufficient to oxidize the reduced ceria sample at RT in short span of time about minutes [190,191]. After t=8 minutes of elapsed irradiation the recovery and reordering by oxygen vacancies filling at the interior region with respect to the stationary crystal-3 is seen. This t=8 minutes recorded HRTEM image is shown in fig.3.24 (c). The physics of this process is similar to the solid-state lattice-mending around nanopores in the case of laser annealing for which the impetus to generate crystalline order comes from the combined effect of pressure and the atomic thermal diffusion efficacy [192–194].

This process is similar to bond-breaking and bond-making phenomenon, which is driven by the defect-induced system energy minimization. Significantly, this was also observed in nanocrystalline CuO nucleation and growth progression by in-situ TEM studies by the present author [51]. In which studied 2D-$Cu_2(OH)_3NO_3$ single-crystalline nanoflakes were used as a representative material. The demonstration of nanocrystalline CuO growth under TEM e-beam is supportive of the present argument. The uniqueness lies in reliving the (differential dislocation densities, crystalline boundary curvature, strain, etc.) energetically different crystalline entities either individually or as a combination which get stimulated healing under the influence of the TEM e-beam fluence in a controlled fashion [51,195–198]. Thus, the oxygen vacancy generation, lattice-mending around these oxygen vacancies and the oxygen vacancies annihilation facilitates the reordering under TEM e-beam as is quite evident from fig.3.24 (c). Subsequent snapshot in the fig.3.24 (d) indicates development of a renewed orderly nanocrystalline region in the previously spotted crystal-2 (see fig.3.24 (a) crystal-2 at t=0 min) having a rotation between <022> direction of approximately 12 °, observed in [01-1] zone axis. Further, the orderly crystalline region extends to crystal-1 region (see fig.3.24 (a) crystal-1 at t=0 min) after a time-lapse of t=16 min snapped in the fig.3.24 (e) as a 14 nm larger subunit, which actually lacks atomic periodicity. Hence, it is missing after contrast enhancement with respect to the background shown in the fig.3.24 (d). The growth front analysis of the 14 nm grown subunit in fig.3.24 (e), is realized. The one to one correspondence of HRTEM lattice fringes and corresponding FFT of the grown crystal suggests that the OA happens in {011} with the probe TEM e-beam traveling in [01-1] as the zone-axis. As stated in the earlier subsection, this TEM e-beam facilitated OA growth will continue until the entire TEM e-beam hammered flattened flake region grows into a single larger subunit.

## TEM e-beam Un-hammered region under e-beam exposure:

A region having crystallites free to move laterally is continually observed under TEM e-beam. Even reordering of an amorphous core of a crystallite with time occurred. However, no attachment of the oriented crystallites leading to growth is observed. Snapshots of all these in-situ traced features are presented in the figs.3.25 (a)-(d). It justifies the importance of the TEM e-beam hammering and preparation of a radiation-hardened region having crystallites to demonstrate the OA mechanism and subsequent growth facilitated under TEM e-beam as the driving agent.

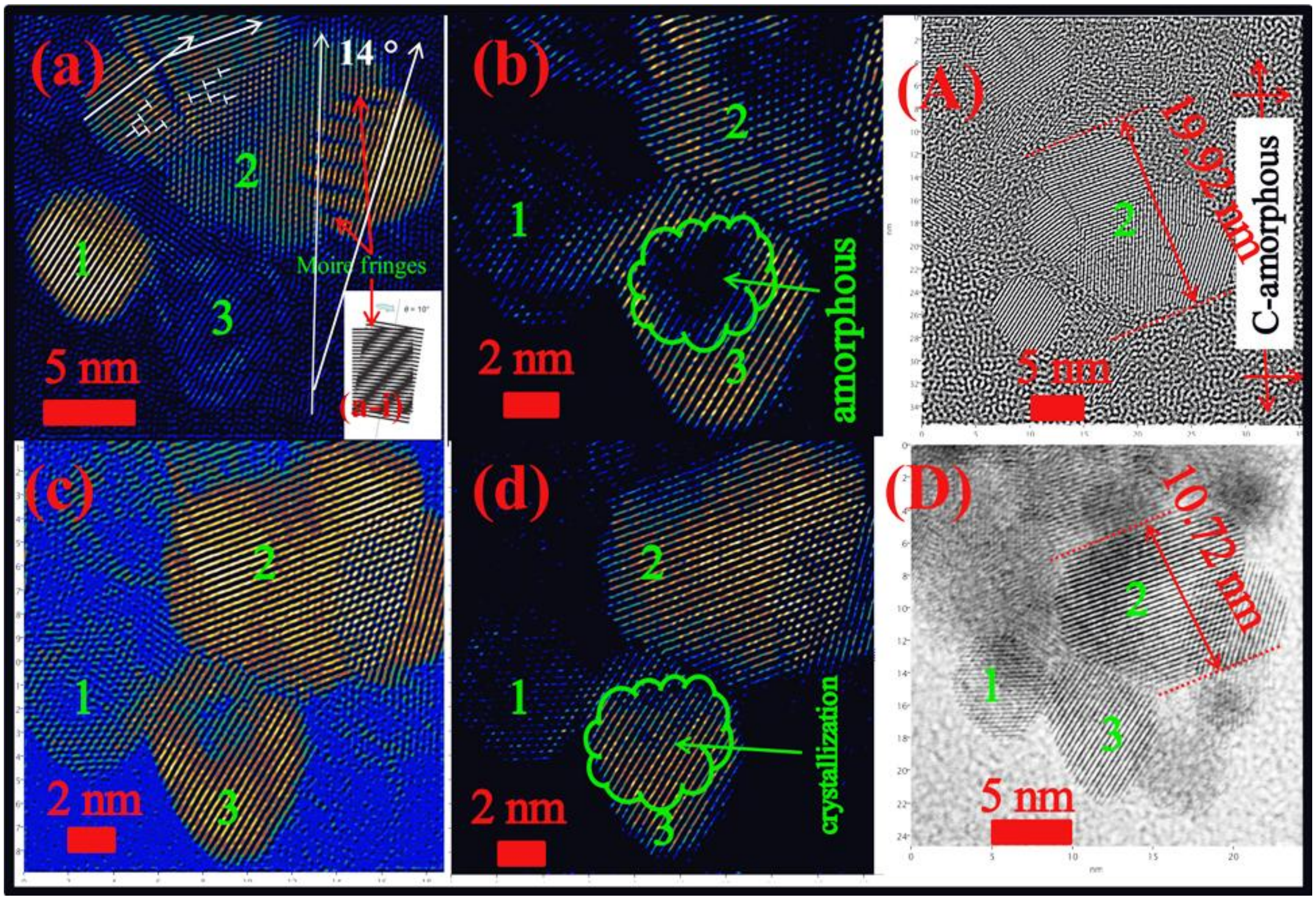

**Fig.3.25** [***Probe Region within which crystallites are free to move laterally***]: Simultaneous reordering and damage of crystallites sequential in-situ HRTEM time evolved BF images after; (a) t=0, (b) t=4, (c) t=8, and (d) t=12 minutes of step-4 TEM e-beam fluence irradiation. Tracked specific nano-crystallite exposed to TEM e-beam that undergoes size reduction.

## Summary:

The review findings are summarized now. These are listed below.

(1) Morphological hierarchy, mesocrystals of intermediate formations and the biominerals' growth pathway progress are motivating in writing this review. Dominantly observed, aquatic-medium biominerals' growth process is mimicked. The aquatic neighbor's participation in achieving biomineralization, whether an active or as an inert medium, is investigated.

(2) Nanoscale ceria, having demonstrated versatility in biological applications, is the prototype material of choice to regenerate aquatic environment observed calcium and silicon compound developed macrostructures by organisms.

(3) Nanocrystalline ceria ambient crystallization is favorable. However, at physiological pH=7.4 sparingly soluble in water. Thereby ultrasonic probe sonication is employed at RT to deliver water-soluble stable, transparent nc-ceria supernatant colloidal dispersion.

(4) The supernatant is stable for a month, and subsequent gradual settling with aging is investigated. A set of settled mass recovered sequentially for 12 months is observed to follow the NCG pathway. Spectroscopic (Raman and UV-Vis optical) analysis done for this sequential settled products demonstrates surface auto-regenerative CT attribute with aging.

(5) However, the instant settling achieved by adding $H_2O_2$ as the oxidant to the nc-ceria supernatant colloidal dispersion had no characteristic NCG pathway signature, i.e., lumpy aggregates.

(6) The presented NCG pathway by utilizing aging at ambient is one of the parameters shown to contribute to mineral growths, whereas in an aquatic medium, several other parameters also contribute (pressure, temperature, chemical species, and aquatic-medium participation itself). Also, the nc-ceria RT crystallization aspect enables bio-mimicking the presented particle-particle attachment scheme without the use of any additional thermal input.

(7) Also, the DI-water direct participation in delivering 1D-ceria fibers is observed.

(8) These ambient NCG protocol grown 1D-ceria fibers is in the hexagonal-$Ce_2O_3$ crystal phase, illustrating growth anisotropy generated cubic to a hexagonal phase transition.

These fibers crystal facets are evolved out of the highly reactive ceria planes will be a candidate for catalysis applications.

(9) The dual role of the TEM e-beam as a material modification and probe tool is demonstrated. Crystal growth is a radiation-induced defect-stimulated defect healing process at RT. It is demonstrated that the near-neighbor environment around the nc-ceria In the TEM chamber can be used effectively to control growth kinetics and, therefore, properties.

## Acknowledgements:

Facilities provided by Advanced Center of Research in High Energy Materials (ACRHEM), Center for Nanotechnology (CFN), School of Physics (SOP), School of Chemistry (SOC), and Engineering Science and Technology (SEST), UOH are acknowledged. Special thanks, to TEM instrument operator Mr. Pankaj patience for doing the measurements as per S K Padhi instructions. Mr. S K Padhi acknowledges DRDO-India for doctoral fellowship.